\documentclass[journal]{IEEEtran}
\IEEEoverridecommandlockouts
\usepackage{cite}
\usepackage{hyperref}
\usepackage{amsmath,amssymb,amsfonts}
\usepackage{amsmath}
\DeclareMathOperator*{\argmax}{argmax}
\usepackage{graphicx}
\usepackage{textcomp}
\usepackage{xcolor}
\usepackage{graphicx}
\usepackage{float}
\usepackage{subfigure}
\usepackage{amsmath}
\usepackage{amsfonts,amssymb}
\usepackage{mathrsfs}
\usepackage{mathtools}
\usepackage{algorithm}
\usepackage{algorithmicx}
\usepackage{algpseudocode}
\usepackage{stfloats}
\usepackage{subfigure}
\usepackage{multirow}
\usepackage{booktabs}

\bibliography{IEEEabrv}
\def\BibTeX{{\rm B\kern-.05em{\sc i\kern-.025em b}\kern-.08em
    T\kern-.1667em\lower.7ex\hbox{E}\kern-.125emX}}
\begin{document}
\title{Massive MIMO Antenna Selection: Asymptotic Upper Capacity Bound and Partial CSI\\
}

\author{Chongjun~Ouyang,
        and~Hongwen~Yang,~\IEEEmembership{Member,~IEEE}
\thanks{M. Ouyang and M. Yang are with the Wireless Theories and Technologies Lab, Beijing University of Posts and Telecommunications, Beijing, 100876 China e-mail: \{DragonAim, yanghong\}@bupt.edu.cn}
}


\maketitle

\begin{abstract}
Antenna selection (AS) is regarded as the key promising technology to reduce hardware cost but keep relatively high spectral efficiency in multi-antenna systems. By selecting a subset of antennas to transceive messages, AS greatly alleviates the requirement on Radio Frequency (RF) chains. This paper studies receive antenna selection in massive multiple-input multiple-output (MIMO) systems. The receiver, equipped with a large-scale antenna array whose size is much larger than that of the transmitter, selects a subset of antennas to receive messages. A low-complexity asymptotic approximated upper capacity bound is derived in the limit of massive MIMO systems over independent and identical distributed (i.i.d.) Rayleigh flat fading channel, assuming that the channel side information (CSI) is only available at the receiver. 
Furthermore, numerical simulations are provided to demonstrate the approximation precision of the asymptotic results and the tightness of the capacity bound.  
Besides the asymptotic analysis of the upper bound, more discussions on the ergodic capacity of the antenna selection systems are exhibited. By defining the number of corresponding rows in the channel matrix as the amount of acquired CSI, the relationship between the achievable channel capacity and the amount of acquired CSI is investigated. Our findings indicate that this relationship approximately follows the Pareto principle, i.e., most of the capacity can be achieved by acquiring a small portion of full CSI. Finally, on the basis of this observed law, an adaptive AS algorithm is proposed, which can achieve most of the transmission rate but requires much less CSI and computation complexity compared to state-of-the-art methods.
\end{abstract}

\begin{IEEEkeywords}
Massive MIMO, receive antenna selection, upper capacity bound, asymptotic theory, Pareto principle, partial CSI.
\end{IEEEkeywords}

\section{Introduction}
\label{section1}
Massive multiple-input multiple-output (MIMO) is considered to be one of the most prospective technologies in the upcoming 5th generation cellular networks (5G) \cite{b38,b21} and millimeter wave (mmWave) communication \cite{b55} since it can dramatically improve the spectral efficiency \cite{b13}. In addition to the gain in spectral efficiency, massive MIMO can greatly improves the transmission security and reliability \cite{b39,b14}. By deploying large-scale antenna array at the transceiver, massive MIMO remarkably promotes the system performance in contrast to the small-scale systems. However, to promise communication, each antenna should be equipped with a Radio Frequency (RF) chain, which will results in expensive cost and high energy consumption of hardware implementation, particularly in massive MIMO systems. To solve this problem, massive MIMO with antenna selection (AS-MIMO) \cite{b2} has gained significant attentions in recent years aiming for design of high-efficiency transmission schemes \cite{b40,b20,b43,b56}.

Antenna selection (AS) is able to alleviate the requirement on RF chains by selecting a subset of antennas to transceive messages. To analytical measure the performance of receive antenna selection (RAS), Molisch et al. proposed the concept of upper capacity bound in \cite{b33} and derived the statistical distribution of the upper bound for different cases. Dually, Sanayei et al. \cite{b30} investigated the distribution of the channel capacity of MIMO systems with transmit antenna selection (TAS) in the limit of high and low Signal to Noise Ratio (SNR). However, many derivations in conventional small-scale AS-MIMO systems can not be directly applied into massive AS-MIMO systems for the large-scale antenna deployment will cause high computation complexity, for example, the upper bound defined in \cite{b33}. To resolve this issue, many other mathematical tools have been utilized in massive MIMO systems, such as the asymptotic theory \cite{b32,b28}.

Asymptotic theory on order statistics can be applied into massive MIMO systems to simplify some derivations or approximate some system performance due to the properties of large dimensionality originated from large-scale antenna arrays. Up to now, several studies have exploited this theory to perform theoretical analysis\cite{b31,b3,b36,b15,b23,b26,b27}. By assuming infinite transmit or receive antennas, \cite{b31} derived the distribution of channel capacity for MIMO systems based on the asymptotic theory, which serves as the principle of massive MIMO technology. In 2018, Y. Gao et al. \cite{b3} moderately simplified the derivations of the upper capacity bound, defined in \cite{b33}, on the basis of the asymptotic theory. Additionally, with this theory, \cite{b36} derived the asymptotic distribution of the channel capacity for AS systems. Moreover, this theory was utilized in \cite{b15} to explore the energy efficiency of the massive AS-MIMO systems. Following similar thought, \cite{b23} derived a high-precision analytical approximated expression of the channel capacity for massive MIMO systems with TAS. Furthermore, \cite{b26} and \cite{b27} extended \cite{b23} to massive Multiple-Input Multiple-Output Multiple-Eavesdropper (MIMOME) channels \cite{b29} to explore the correlation between the number of RF chains and transmission security.

It is well-known that the acquisition of channel side information (CSI) is significant to antenna selection. Most of the studies have assumed that perfect and full CSI is available by the channel reciprocity\cite{b58}. However, it is hard to obtain the instantaneous channel state or guarantee its correctness either in FDD \cite{b59} or TDD \cite{b60} systems for massive MIMO, since the channel matrix ${\bf{H}}$ is of large-size and the number of users are usually overwhelming. \cite{b22,b34,b44} analyzed the performance of AS-MIMO systems under the influence of imperfect channel side information (CSI). Furthermore, the scenario with partial CSI was discussed in \cite{b45,b48}. \cite{b45} assumed only eigenvectors of ${\bf{H}}^{\dagger}{\bf{H}}$ was available and proposed the optimum transmission strategy. Additionally, in \cite{b48}, only large-scale CSI was known at the transmitter side and a joint antenna selection and beamforming algorithm was formulated to minimize the power consumption.

Besides the performance analysis, many suboptimal capacity-orientated algorithms for antenna selection have been proposed in the conventional MIMO systems \cite{b51,b41,b12,b16,b19}. A correlation-based method was investigated in \cite{b51}, which is fast but suboptimal. The greedy search (GS) were proposed in \cite{b41} and \cite{b12}, respectively. Even though the algorithms shown in \cite{b41,b12} have different forms, they posses the same essence. In addition, \cite{b16} introduced a sub-optimal selection algorithm from the view of convex optimization. Furthermore, a machine learning based selection method was shown in \cite{b19}. 

Even though most of these aforementioned algorithms were first proposed for small-scale MIMO systems, but they have been extended into the massive MIMO systems. On the basis of the correlation-based method in \cite{b51}, a two-step selection algorithm was presented in \cite{b25} which required very low complexity. Additionally, \cite{b11} explored the greedy search in massive MIMO and provided theoretical basis for its near-optimality. The convex optimization was applied into antenna selection for large-scale MIMO systems in \cite{b17,b18,b50} and two sub-optimal algorithms were formulated. \cite{b37} detailedly investigated the norm-based transmit antenna selection strategy, which is of very low-complexity but achieves a relatively small channel capacity in contrast to most selection methods. Moreover, \cite{b53,b54} put forward a suboptimal selection strategy by using the theory of rectangular maximum-volume (RMV) submatrices. Furthermore, an optimal selection algorithm based on branch-and-bound search was shown in \cite{b3}, which can find the optimal antenna subset with much lower complexity  than the exhaustive search. Nevertheless, all of these algorithms require full CSI before the subset search which may consume much resource, such as time and energy. Therefore, it is necessary to explore a new selection algorithm which only requires a small portion of the full CSI but can achieve most of the channel capacity. 

\subsection*{Contribution and Organization}
In this paper, we detailedly investigate the receive antenna selection in massive MIMO systems. 
Following the thought in \cite{b33} and \cite{b3}, this paper still utilizes the defined upper bound to analytical measure the AS-MIMO systems and a new calculation method is introduced. 
As stated before, it is extremely hard to acquire full even perfect CSI. Since many studies have investigated the scenarios of imperfect CSI, this article focus on the scenario when only partial CSI is acquired. In addition, an adaptive antenna selection algorithm is proposed by using only partial CSI. 
Taken together, the key contributions of this paper can be summarized as follows:
\begin{itemize}
\item
To lower the computation complexity, asymptotic theory is used to approximate the upper channel capacity due to the property of high-dimensionality of massive MIMO. By the definition of upper capacity bound in \cite{b33}, the asymptotic approximation is discussed in two scenarios: 1) For Scenario A: the number of the selected antennas is no larger than that of the transmit antennas, and 2) For Scenario B: the amount of the selected antennas exceed that of the transmit antennas. In each scenario, simulation results demonstrate that the derived asymptotic bound has good approximation effect. Compared with the approximation strategy used in \cite{b3}, our derivation holds a much lower complexity.   
\item
Channel side information is vital for antenna selection and this paper defines the number of rows in the channel matrix as the amount of CSI. Furthermore, taking CSI acquisition into consideration, we investigate the correlation between the achievable efficient capacity, defined in \cite{b3}, and the number of acquired CSI. Our investigation shows that there exists an optimal number of CSI for antenna selection to achieve the largest efficient capacity.    
\item
The relationship between the achievable channel capacity and the amount of acquired CSI is explored in detail. Our findings suggest that this relationship follows the Pareto principle, i.e., most of the transmission rate can be achieved by acquiring a small portion of full CSI. By this observed law, an adaptive algorithm is designed which only requires partial CSI. Compared with the state-of-the-art AS methods, the proposed algorithm holds a much lower complexity with the guarantee of high achievable channel capacity.    
\end{itemize}

The remaining parts of this paper is structured as follows. Section \ref{sec2} describes the system model and presents the problems to be settled in this paper. In Section \ref{sec3}, the asymptotic theory is utilized to approximate the upper bound of the massive AS-MIMO systems for both BUB and MUB scenarios. More detailed discussions about the ergodic capacity of AS-MIMO systems are exhibited in Section \ref{sec4}. In Section \ref{sec5}, an adaptive selection algorithm is proposed on the basis of the explored inherent characteristics of the massive AS-MIMO systems. Finally, Section \ref{sec6} concludes the paper.

$Notations$: Throughout this paper, scalars, vectors and matrices are denoted by non-bold, bold lower case, and bold upper letters, respectively. $\mathbb{C}$ stands for the complex numbers. The Hermitian and inverse of matrix $\bf{H}$ is indicated with ${\bf{H}}^{\dagger}$ and ${\bf{H}}^{-1}$, and ${\bf{I}}_N$ is the $N{\times}N$ identity matrix. $\det(\cdot)$ and ${\bf{E}}(\cdot)$ denote the determinant and expectation operator. Moreover, the complex and real Gaussian distribution with mean $\eta$ and variance $\sigma^2$ are separately represented by ${\mathcal{CN}}(\eta,\sigma^2)$ and ${\mathcal{N}}(\eta,\sigma^2)$.

\section{System Model and Problem Formulation}
\label{sec2}
In this paper, we consider a massive MIMO system in which the transmitter is equipped with $N_{\rm{t}}$ antennas and the receiver is equipped with $N_{\rm{r}}$ antennas ($N_{\rm{r}}\gg N_{\rm{t}}$). At the receiver, only $L$ ($L\ll N_{\rm{r}}$) RF chains are deployed, thus the receiver has to select $L$ antennas to receive messages.
\subsection{System Model}
In the aforementioned massive MIMO system, the received signal vector is
\begin{equation}
{{\bf{y}}}=\sqrt{{\rho}}{{\bf{H}}}\bf{x}+{{\bf{w}}},
\end{equation}
where ${\bf{x}}\in{\mathbb{C}}^{N_{\rm{t}}\times1}$ is the transmitted signal, ${{\rho}}$ is the SNR at each receive antenna, and ${{\bf{w}}}{\sim}{\mathcal{CN}}{({\bf{0}},{{\bf{I}}_{N_{\rm{r}}}})}$ is the additive complex Gaussian noise. Moreover, suppose that the transmitted signals from different antennas are independent and ${\bf{E}}\left[{\bf{x}}^{\dagger}{\bf{x}}\right]=1$. Considering independent and identically distributed (i.i.d) Rayleigh flat fading channels, all the elements in the channel matrix ${\bf{H}}{\in}{{\mathbb{C}}^{{N_{\rm{r}}}{\times}{N_{\rm{t}}}}}$ follow ${\mathcal{CN}}(0,1)$. Furthermore, assume that the channel side information is only available at the receiver and the transmit power is uniformly allocated, thus the channel capacity can be written as \cite{b1}
\begin{equation}
{C}=\log_2\det\left({{\bf{I}}_{N_{\rm{r}}}}+\frac{{\rho}}{N_{\rm{t}}}{\bf{H}}{{\bf{H}}^{\dagger}}\right).
\end{equation}

As stated before, $L$ antennas are selected or activated for communication at the receive end. 
The channel capacity after receive antenna selection is given by
\begin{equation}
{\tilde{C}}=\log_2\det\left({{\bf{I}}_{L}}+{\overline{\rho}}{\tilde{\bf{H}}}{{\tilde{\bf{H}}}}^{\dagger}\right),
\end{equation}
where ${\overline{\rho}}=\frac{{\rho}}{N_{\rm{t}}}$ is defined as the normalized SNR and ${{\tilde{\bf{H}}}}{\in}{{\mathbb{C}}^{{L}{\times}{N_{\rm{t}}}}}$ denotes the submatrix after RAS. Let $\mathcal{T}$ denote the selected subset of receive antenna indexes whose cardinality is $\left|{\mathcal{T}}\right|=L$. In terms of the channel capacity, the RAS problem is formulated as
\begin{equation}
{\mathcal{T}}^{\rm{opt}}=\argmax_{{\mathcal{T}}{\in}{\mathcal{M}}}\log_2\det\left({{\bf{I}}_{L}}+{\overline{\rho}}{\tilde{\bf{H}}}{{\tilde{\bf{H}}}}^{\dagger}\right),
\label{EQU4}
\end{equation}
where $\mathcal{M}$ denotes the full set of all the candidate antenna index subsets with size $L$. Define $C_{\rm{s}}$ as the channel capacity of the AS-MIMO system and let ${\tilde{\bf{H}}}_{\rm{s}}$ denote the corresponding submatrix of ${\mathcal{T}}^{\rm{opt}}$, thus $C_{\rm{s}}$ can be given by
\begin{equation}
{C_{\rm{s}}}=\log_2\det\left({{\bf{I}}_{L}}+{\overline{\rho}}{\tilde{\bf{H}}}_{\rm{s}}{\tilde{\bf{H}}}{_{\rm{s}}^{\dagger}}\right).
\label{EQU5}
\end{equation}
 
\subsection{Efficient Channel Capacity}
It is necessary to acquire the CSI in order to do antenna selection at the receiver. In this paper, the number of rows in the channel matrix $\bf{H}$ is used to measure the amount of acquired CSI, which is denoted by $\Upsilon$. For example, in receive antenna selection, $\Upsilon=20$ means that only 20 rows of the channel matrix are acquired and utilized during the RAS. Moreover, an efficient channel capacity \cite{b3} can be defined in terms of CSI acquisition. Let $t_{\rm{tr}}$ and $t_{\rm{coh}}$ denote the training duration and coherence time, respectively. Since only $L$ RF chains are available at the receiver, they must be used for multiple rounds to obtain enough CSI. Therefore, the time for full CSI acquisition is $\frac{N_{\rm{r}}}{L}t_{\rm{tr}}$. As a result, when full CSI is acquired during the RAS, the efficient channel capacity can be calculated as following:
\begin{equation}
\begin{split}
{C_{\rm{s,eff}}}&={C_{\rm{s}}}\frac{t_{\rm{coh}}-{N_{\rm{r}}}/{L}t_{\rm{tr}}}{t_{\rm{coh}}}={C_{\rm{s}}}\left(1-\frac{N_{\rm{r}}}{L}\frac{t_{\rm{tr}}}{t_{\rm{coh}}}\right)\\
&={C_{\rm{s}}}\left(1-\frac{N_{\rm{r}}}{L}\eta\right),
\end{split}
\end{equation}
where $\eta=\frac{t_{\rm{tr}}}{t_{\rm{coh}}}$ and it is apparent that smaller $\eta$ indicates higher acquisition efficiency. 

Actually, the definition of the efficient capacity has many limitations. For example, the time used for antenna selection is not taken into consideration when calculating the efficient capacity. For the sake of brevity, this paper will focus more on the exact ergodic capacity without considering the time consumption originated from CSI acquisition and antenna selection.    

\subsection{Upper Bound}
The channel capacity after RAS is determined by \eqref{EQU5}, whereas it seems difficult to utilize this formula for its prohibitive computation complexity stemming from exhaustive search (ES). Therefore, it makes sense to define the capacity upper bound to measure the performance of antenna selection technology. However, the computation complexity is still huge in order to obtain the analytical form of the upper bound, defined in \cite{b33}, especially in massive MIMO systems. Even if the computation complexity has dropped in \cite{b3}, it is still unacceptable in large-scale systems. As a result, here comes the first question: does a more advanced and efficient solution to the upper bound exist? 

\subsection{Antenna Selection Algorithm} 
As stated before, CSI acquisition is vital to antenna selection. However, CSI acquisition will consume many resources, such as and time and energy, which may lead to the degradation of spectral and energy efficiency. Most of the antenna selection algorithms require full CSI and this is definitely a great burden for the large-scale system. Additionally, the antenna selection problem defined in \eqref{EQU4} is NP-hard \cite{b52}which may result in exponential complexity since there exists no high-efficient selection strategies. These challenges lead to the second problem: is there a selection algorithm which requires only partial CSI and much less complexity with the guarantee of relatively high achievable channel capacity?

\section{Asymptotic Approximated Upper Bound}
\label{sec3}
Molisch et al. \cite{b33} defined two types of upper capacity bound to analytically measure the performance of antenna selection in MIMO systems and these definitions are still effective for massive MIMO. The first type of the upper bound is utilized in the scenario when $L{\leq}N_{\rm{t}}$. In this case, the entire system is regarded as $N_{\rm{r}}$ independent multiple-input single-output (MISO) subsystems. Transmit beamforming (BF) is used in each subsystem and the best $L$ ones of these subsystems are selected. As for the second type, it is used when $L{>}N_{\rm{t}}$. More specifically, the whole system is treated as $N_{\rm{t}}$ independent single-input multiple-output (SIMO) subsystems and the best $L$ receive antennas are activated for maximal ratio combination (MRC) in each subsystem. Throughout this paper, the term BF Upper Bound (BUB) and MRC Upper Bound (MUB) are used to refer to these two bounds respectively. In each scenario, suppose that full CSI is utilized at the receiver i.e., $\Upsilon=N_{\rm{r}}$. 

\subsection{BF Upper Bound}
\label{sec4.1}
Foschini and Gans first proposed the concept of the upper capacity bound for the full-complexity MIMO system in \cite{b1}, which is given by
\begin{equation}
\label{equ5}
{C_{\rm{full}}}=\sum_{i=1}^{N_{\rm{t}}}\log_2\left(1+{\overline{\rho}}\alpha_i\right),
\end{equation}
where $\{\alpha_i\}_{i=1,2,\cdots,N_{\rm{t}}}$ are i.i.d chi-square random variables with $2N_{\rm{r}}$ degrees of freedom. A virtual situation is displayed in \eqref{equ5} where each of the $N_{\rm{t}}$ transmitted signals is received by a separate set of $N_{\rm{r}}$ receive antennas without interference from each other\cite{b1}. Similarly, a new upper capacity bound can be defined by exchanging the role of the transmitter and receiver, which is written as
\begin{equation}
\label{equ6}
{{\tilde{C}}_{\rm{full}}}=\sum_{i=1}^{N_{\rm{r}}}\log_2\left(1+{\overline{\rho}}\gamma_i\right),
\end{equation}
where $\{\gamma_i\}_{i=1,2,\cdots,N_{\rm{r}}}$ are i.i.d chi-square random variables with $2N_{\rm{t}}$ degrees of freedom. This new definition still indicates an artificial case when each of the $N_{\rm{r}}$ receive antennas has its own subset of transmit antennas without considering the interference among these subsystems. On the basis of \eqref{equ6}, the upper capacity bound for RAS can be written as follows:
\begin{equation}
\label{equ7}
{{\tilde{C}}_{\rm{s}}}=\sum_{i=1}^{L}\log_2\left(1+{\overline{\rho}}\gamma_{(i)}\right),
\end{equation}
where $\{\gamma_{(i)}\}_{i=1,2,\cdots,N_{\rm{r}}}$ ($\gamma_{(1)}{\geq}\gamma_{(2)}{\geq}{\cdots}{\geq}\gamma_{(N_{\rm{r}})}$) are $ordered$ chi-squared-distributed variables with $2N_{\rm{t}}$ degrees of freedom.
This bound has been proved to be relatively tight when $L{\leq}N_{\rm{t}}$ holds\cite{b33}. Furthermore, the smaller $L$ is, the tighter the upper bound is \cite{b33}. Considering the extreme case when $L=1$ in a manner where the MIMO system after RAS degrades into a MISO system, then the upper bound just equals to the channel capacity $C_{\rm{s}}$ in \eqref{EQU5}. Nevertheless, the calculation of the joint distribution of the top-$L$ $ordered$ statistics from $\left\{\log_2\left(1+{\overline{\rho}}\gamma_{(i)}\right)\right\}_{i=1,2,\cdots,N_{\rm{r}}}$ is computationally complex, especially when $N_{\rm{r}}$ is large\cite{b33,b3}. Therefore, it is vital to explore a low-complexity method to calculate the upper bound.

In the sense of large-scale behavior, the asymptotic theory has become a topic of interest to alleviate computation complexity. With the guarantee of high precision, the asymptotic theory derives an approximate distribution of the top-$L$ variables instead of calculating the exact joint distribution of them. 
In asymptotic theory, $\sum\limits_{i=1}^{L}\log_2\left(1+{\overline{\rho}}\gamma_{(i)}\right)$ is termed as a trimmed sum\cite{b4}. With the total size $N_{\rm{r}}$ tending to infinity, the trimmed sum is shown to converge to a Gaussian random variable\cite{b4}. On the other hand, simulation results indicate that the distribution of $\sum\limits_{i=1}^{L}\log_2\left(1+{\overline{\rho}}\gamma_{(i)}\right)$  draws fast convergence rate with the increment of $N_{\rm{r}}$. Therefore, a normal approximation can be applied to the trimmed sum even though the range size $N_{\rm{r}}$ is of limited length. 
Based on the main theorem in \cite{b4}, $\sum\limits_{i=1}^{L}\log_2\left(1+{\overline{\rho}}\gamma_{(i)}\right)$ can be approximated as a Gaussian random variable $g{\sim}{{\mathcal{N}}\left(\mu_g,\sigma_g^2\right)}$, in which $\mu_g$ and $\sigma_g^2$ are determined as
\begin{subequations}
\label{equ8}
\begin{align}
{\mu_g}&=N_{\rm{r}}\int_u^{\infty}\log_2\left(1+{\overline{\rho}}x\right){f_{N_{\rm{t}}}\left(x\right)}{\rm{d}}x\\
{\sigma_g^2}&=L\left(\sigma^2+\left(u-\frac{\mu_g}{L}\right)^2\left(1-\frac{L}{N_{\rm{r}}}\right)\right),
\end{align}
\end{subequations}
where
\begin{equation}
{\sigma^2}=\frac{N_{\rm{r}}}{L}\int_u^{\infty}\left(\log_2\left(1+{\overline{\rho}}x\right)\right)^2{f_{N_{\rm{t}}}\left(x\right)}{\rm{d}}x-\frac{\mu{_g^2}}{L^2},
\label{sigma}
\end{equation}
and ${f_{N_{\rm{t}}}\left(\cdot\right)}$ denotes the chi-squared probability density function (PDF) with $2N_{\rm{t}}$ degrees of freedom and mean $N_{\rm{t}}$ which reads\cite{b8}
\begin{equation}
{f_{N_{\rm{t}}}\left(x\right)}=\frac{1}{\left({N_{\rm{t}}}-1\right)!}\left\{\begin{aligned}&{\rm{e}}^{-x}x^{N_{\rm{t}}-1},&x{\geq}0\\&0,&x{<}0\end{aligned}\right..
\label{equ9}
\end{equation}
The constant $u$ in \eqref{equ8} satisfies $\int_u^{\infty}{f_{N_{\rm{t}}}\left(x\right)}{\rm{d}}x=\frac{L}{N_{\rm{r}}}$ which can be easily solved with the {\bf{MATLAB}} function $chi2inv(\cdot,\cdot)$. By substituting \eqref{equ9} into \eqref{equ8} and \eqref{sigma}, $\mu_g$ and $\sigma^2$ can be simplified after some derivations, which are exhibited on the top of the next page.  Due to the attenuation of the term ${\rm{e}}^{-x}\left(1+{\overline{\rho}}x\right)^{-1}$, the integrals in \eqref{equ10a} and \eqref{equ10b} can be solved efficiently by numerical integration.
\newcounter{mytempeqncnt}
\begin{figure*}[!t]
\normalsize
\setcounter{mytempeqncnt}{\value{equation}}
\setcounter{equation}{12}
\begin{subequations}
\begin{align}
{\mu_g}&=N_{\rm{r}}\int_u^{\infty}\log_2\left(1+{\overline{\rho}}x\right){f_{N_{\rm{t}}}\left(x\right)}{\rm{d}}x=N_{\rm{r}}\int_u^{\infty}\log_2\left(1+{\overline{\rho}}x\right){\rm{d}}\left(-\sum_{k=0}^{N_{\rm{t}}-1}\frac{x^k}{{\rm{e}}^xk!}\right)\notag\\&=\frac{{N_{\rm{r}}}}{\ln2}\sum_{k=0}^{N_{\rm{t}}-1}\ln\left(1+{\overline{\rho}}u\right)\frac{u^k}{{\rm{e}}^uk!}+\frac{\overline{\rho}}{\ln2}\sum_{k=0}^{N_{\rm{t}}-1}\int_{u}^{\infty}\frac{{N_{\rm{r}}}x^k}{{\rm{e}}^x\left(1+{\overline{\rho}}x\right)k!}{\rm{d}}x \label{equ10a}\\
{\sigma^2}&=N_{\rm{r}}\int_u^{\infty}\left(\log_2\left(1+{\overline{\rho}}x\right)\right)^2{f_{N_{\rm{t}}}\left(x\right)}{\rm{d}}x-\frac{\mu{_g^2}}{L^2}=N_{\rm{r}}\int_u^{\infty}\left(\log_2\left(1+{\overline{\rho}}x\right)\right)^2{\rm{d}}\left(-\sum_{k=0}^{N_{\rm{t}}-1}\frac{x^k}{{\rm{e}}^xk!}\right)-\frac{\mu{_g^2}}{L^2}\notag\\&=\sum_{k=0}^{N_{\rm{t}}-1}{N_{\rm{r}}}\left(\log_2\left(1+{\overline{\rho}}u\right)\right)^{2}\frac{u^k}{{\rm{e}}^uk!}+\frac{2\overline{\rho}}{\ln2}\sum_{k=0}^{N_{\rm{t}}-1}\int_{u}^{\infty}\frac{{N_{\rm{r}}}x^k\log_2\left(1+{\overline{\rho}}x\right)}{{\rm{e}}^x\left(1+{\overline{\rho}}x\right)k!}{\rm{d}}x-\frac{\mu{_g^2}}{L^2}.\label{equ10b}
\end{align}
\label{equ10}
\end{subequations}
\setcounter{equation}{\value{mytempeqncnt}}
\hrulefill
\vspace*{4pt}
\end{figure*}

\begin{figure}[!t] 
\setlength{\abovecaptionskip}{-5pt} 
\centering 
\includegraphics[width=0.45\textwidth]{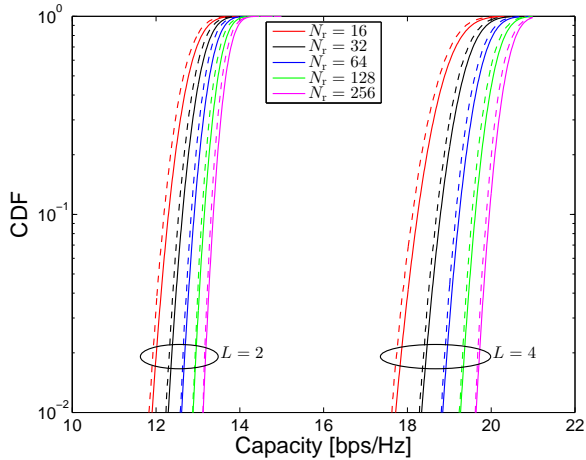} 
\caption{CDF of the asymptotic approximated BF upper bound and the exact BF upper bound, ${\overline{\rho}}=8{\text{dB}}$ and $N_{\rm{t}}=8$. The solid and dashed lines indicate the asymptotic approximated and exact distribution, respectively.} 
\label{figureb1}
\end{figure}

To verify the approximation precision, {\figurename} \ref{figureb1} compares the asymptotic approximated and empirical distribution of the upper capacity bound. As indicated before, ${{\tilde{C}}_{\rm{s}}}$ is approximated with a Gaussian random variable, whose mean and variance are calculated by \eqref{equ10}. The empirical distribution is obtained by Monte-Carlo simulation. It can be seen from this figure that the approximated results nearly coincidence with the simulated results. Moreover, the approximation effect is still satisfying even though $N_{\rm{r}}$ is small, such as 32. In summary, it makes sense to apply asymptotic theory to the approximation of upper capacity bound. 

\begin{figure}[!t]
\setlength{\abovecaptionskip}{-5pt}
\centering
\includegraphics[width=0.45\textwidth]{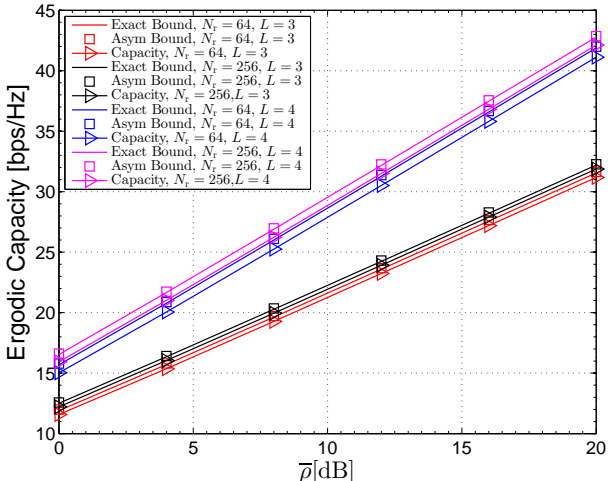}
\caption{Ergodic capacity versus ${\overline{\rho}}$ when ${L}{\leq}{N_{\rm{t}}}$, $N_{\rm{t}}=8$. Asymptotic approximated bound, exact bound and channel capacity $C_{\rm{s}}$ are denoted by Asym Bound, Exact Bound and Capacity, respectively.}
\label{figure1}
\end{figure}

{\figurename} \ref{figure1} compares the ergodic values of the asymptotic approximated and exact upper capacity bound, which are obtained by \eqref{equ10a} and computer simulation, respectively. As is shown in this graph, the approximated and exact results are nearly equal, which further supports the previous derivations. To investigate the tightness of the (asymptotic approximated) upper bound, the exact ergodic capacity by exhaustive search is also presented. According to the simulation results, the BF upper bound is extremely tight. In addition, the bound becomes loose when $L$ increases from 3 to 4, which is consistent with the previous discussions.       

Thus far, this section has investigated the upper bound when $L\leq N_{\rm{t}}$. Nevertheless, the upper capacity bound for the AS-MIMO should be rewritten when the number of activated antennas is larger than  $N_{\rm{t}}$.

\subsection{MRC Upper Bound}
\label{sec4.2}
As for the case when ${L}{>}{N_{\rm{t}}}$, the upper capacity bound is defined as \cite{b33}
\begin{equation}
{\tilde{C}}_{\rm{s}}=\sum_{h=1}^{N_{\rm{t}}}\log_2\left(1+{\overline{\rho}}\sum_{i=1}^{L}{\tilde{\gamma}_{(i)}}\right)=\sum_{h=1}^{N_{\rm{t}}}\xi_h,
\label{equ11}
\end{equation}
where $\{\tilde{\gamma}_{(i)}\}_{i=1,2,\cdots,N_{\rm{r}}}$ ($\tilde{\gamma}_{(1)}{\geq}\tilde{\gamma}_{(2)}{\geq}{\cdots}{\geq}\tilde{\gamma}_{(N_{\rm{r}})}$) are $ordered$ chi-squared-distributed variables with 2 degrees of freedom. An unrealistic scenario is presented in \eqref{equ11} when each of the $N_{\rm{t}}$ antennas communicates with a separate receive antenna subsets with size $N_{\rm{r}}$ in a manner where no interferences among these independent SIMO subsystems occur\cite{b33}. The best $L$ receive antennas are activated for maximal ratio combination in each subsystem, which is also referred to hybrid
selection/maximum ratio combining (H-S/MRC) \cite{b5,b6}. 
Furthermore, Molisch et al. \cite{b33} has proved that this bound is relatively tight especially when $L$ is large.  

On the basis of asymptotic theory, the trimmed sum $\sum_{i=1}^{L}{\tilde{\gamma}_{(i)}}$ can be approximated as a Gaussian random variable $t\sim\mathcal{N}\left(\mu_t,\sigma_t^2\right)$ with mean and variance given by
\begin{subequations}
\label{equ12}
\begin{align}
{\mu_t}&=N_{\rm{r}}\int_u^{\infty}x{f_{1}\left(x\right)}{\rm{d}}x\\
{\sigma_t^2}&=L\left(\sigma^2+\left(u-\frac{\mu_t}{L}\right)^2\left(1-\frac{L}{N_{\rm{r}}}\right)\right),
\end{align}
\end{subequations}
where
\begin{equation}
{\sigma^2}=\frac{N_{\rm{r}}}{L}\int_u^{\infty}x^2{f_{1}\left(x\right)}{\rm{d}}x-\frac{\mu{_t^2}}{L^2},
\end{equation}
and  $f_1(x)={\rm{e}}^{-x}$ denotes the PDF of $\tilde{\gamma}_{(i)}$. The constant $u$ satisfies $\int_{u}^{\infty}f_1(x){\rm{d}}x=\frac{L}{N_{\rm{r}}}$, and thus $u=\ln\frac{N_{\rm{r}}}{L}$. After substitutions and simplifications, ${\mu_t}$ and ${\sigma_t^2}$ can be simplified as follows:
\begin{subequations}
\label{equ14}
\begin{align}
{\mu_t}&=L\left(1+\ln\frac{N_{\rm{r}}}{L}\right)\\
{\sigma_t^2}&=L\left(2-\frac{L}{N_{\rm{r}}}\right).
\end{align}
\end{subequations}
The approximation precision has been investigated in \cite{b36}, which indicates the Gaussian approximation of the trimmed sum $\sum_{i=1}^{L}{\tilde{\gamma}_{(i)}}$ holds a remarkable approximation effect. 

Let $\mu_x$ and $\sigma_x^2$ denote the mean and variance of the asymptotic approximated bound. Since $\{\xi_h\}_{h=1,2,\cdots,N_{\rm{t}}}$ in \eqref{equ11} are i.i.d random variables, $\mu_x$ and $\sigma_x^2$ can be given by
\begin{subequations}
\begin{align}
{\mu_x}&=N_{\rm{t}}\int_0^\infty{\frac{\log_2\left(1+{\overline{\rho}}x\right)}{\sqrt{2\pi\sigma_t^2}}{\rm{e}}^{-\frac{\left(x-\mu_t\right)^2}{2\sigma_t^2}}{\rm{d}}x}\label{mx}\\
{\sigma_x^2}&=N_{\rm{t}}\left(\int_0^\infty{\frac{\left(\log_2\left(1+{\overline{\rho}}x\right)\right)^2}{\sqrt{2\pi\sigma_t^2}}{\rm{e}}^{-\frac{\left(x-\mu_t\right)^2}{2\sigma_t^2}}{\rm{d}}x}-\mu{_x^2}\right).\label{vx}
\end{align}
\end{subequations}

\begin{figure}[!t]
\setlength{\abovecaptionskip}{-5pt}
\centering
\includegraphics[width=0.45\textwidth]{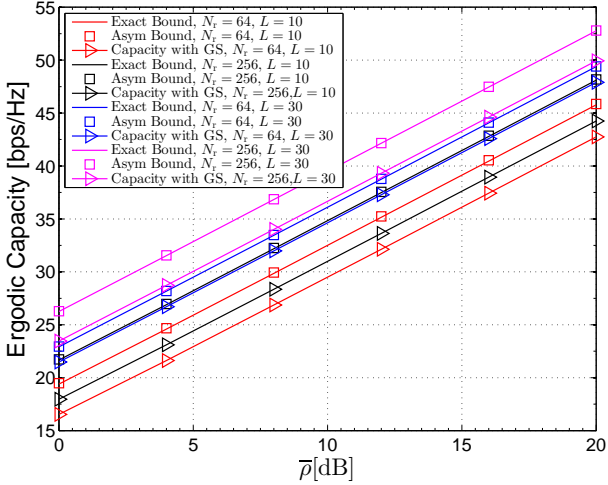}
\caption{Ergodic capacity versus ${\overline{\rho}}$ when ${L}{>}{N_{\rm{t}}}$, $N_{\rm{t}}=4$. Asymptotic approximated bound, exact bound and channel capacity of greedy search are denoted by Asym Bound, Exact Bound and Capacity with GS, respectively.}
\label{figure2}
\end{figure}

{\figurename} \ref{figure2} presents the ergodic capacity versus $\tilde{\rho}$ for different set-up. The asymptotic approximated upper bound and the exact bound are obtained by \eqref{mx} and Monte-Carlo simulation, respectively. It is apparent from {\figurename} \ref{figure2} that there is no clear difference between the asymptotic approximated and exact results, which reflects the approximation accuracy. Due to the huge complexity of exhaustive search when $L$ is large, it is hard to obtain the exact channel capacity $C_{\rm{s}}$. However, greedy search can achieve near-optimal performance which reaches above 90\% of the optimal value according to the work in \cite{b9,b10,b11}, thus it can serve as a benchmark for antenna selection instead. As {\figurename} \ref{figure2} shows, curves representing the greedy search and the upper bound are close. On the other hand, the exact channel capacity is between the bound and the greedy search based channel capacity. Therefore, the upper bound is also close with the exact channel capacity. Moreover, it can be seen from this figure that the upper bound becomes tighter when $L$ gets larger, which is consistent with the former statements. 

\section{Further Explorations Based On The Asymptotic Upper Bound}
\label{sec4}
\begin{figure}[!t]
\setlength{\abovecaptionskip}{-5pt}
\centering
\includegraphics[width=0.45\textwidth]{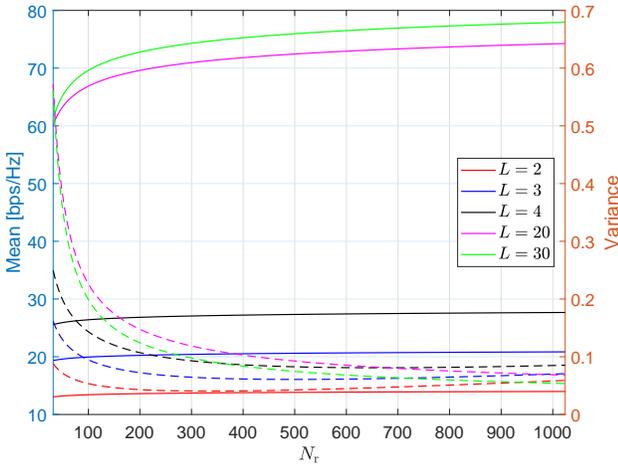}
\caption{Mean and variance of asymptotic approximated upper bound versus $N_{\rm{r}}$, $N_{\rm{t}}=8$ and ${\overline{\rho}}=8{\text{dB}}$. The solid and dashed lines denote the mean and variance, respectively.}
\label{figure3}
\end{figure}

The basic ideas of the asymptotic approximated upper bound has been detailedly investigated in the previous section. This chapter will continue to explore the ergodic capacity of the massive AS-MIMO systems on the basis of the previous derivations. {\figurename} \ref{figure3} illustrates the mean and variance of the asymptotic approximated bound. It is shown that the variance converges to a tiny value gradually as $N_{\rm{r}}$ increases, which indicates that the bound will become more concentrated. The mean value gradually stabilizes or increases slowly as  $N_{\rm{r}}$ increases. For the BF upper bound, there is even no clear difference when $N_{\rm{r}}$ equals to a very small or large value. These can be treated as the results of channel hardening effect \cite{b15,b31}. Such effects will be much more highlighted if the performance criterion is replaced with the exact channel capacity.

It has been investigated before that the upper bound is relatively tight. To precisely estimate the ergodic capacity for the AS-MIMO systems, more work needs to be done. Let $\bar{C}_{\rm{s}}$ and $\bar{C}_{\rm{asy}}$ denote the exact ergodic capacity and the mean value of the asymptotic upper bound, respectively. Therefore,
\begin{equation}
\label{equ17}
\bar{C}_{\rm{s}}=\bar{C}_{\rm{asy}}-\Xi,
\end{equation}
where $\Xi$ represents the gap between the exact value and the asymptotic approximated bound. For the BF upper bound, the gap $\Xi$ is approximated as follows
\begin{equation}
\Xi\approx\frac{{0.1146L^2}(L-1)}{{N_{\rm{t}}^{0.4401\sqrt{L}}}}\left\{\begin{aligned}&1,&\tilde{\rho}\geq0~ {\rm{dB}}\\&\frac{{\rm{e}}^{0.2226({\tilde{\rho}}+8.78)}}{{\rm{e}}^{0.2226({\tilde{\rho}}+8.78)}+1},&\tilde{\rho}<0~ {\rm{dB}}\end{aligned}\right.,
\label{equ18}
\end{equation}
where the normalized SNR is expressed in dB i.e., $10\lg(\cdot)$. This approximation is based on lots of simulation experiments and can be treated as a empirical formula.\footnote{Actually, the approximation of $\Xi$ is still an open problem. More work can be done at this point and we only offer a rough approximation result. Simulation results in the following parts show that this approximation works well.} Since the approximation for the gap $\Xi$ is not the key point of this article, more detailed demonstrations for this formula are not exhibited here. In summary, the approximated ergodic capacity when $L\leq{N_{\rm{t}}}$ can be given by
\begin{equation}
\bar{C}_{\rm{s}}\approx\bar{C}_{\rm{asy}}+\tilde{\Xi},
\label{equ20}
\end{equation} 
where $\tilde{\Xi}$ is the approximation for $\Xi$ shown in \eqref{equ18}. Nevertheless, for the case when $L>{N_{\rm{t}}}$, it is difficult to simulate the ergodic capacity due to the high computation complexity, which makes it even harder to do the approximation.

\begin{figure}[!t] 
\setlength{\abovecaptionskip}{-5pt} 
\centering 
\includegraphics[width=0.45\textwidth]{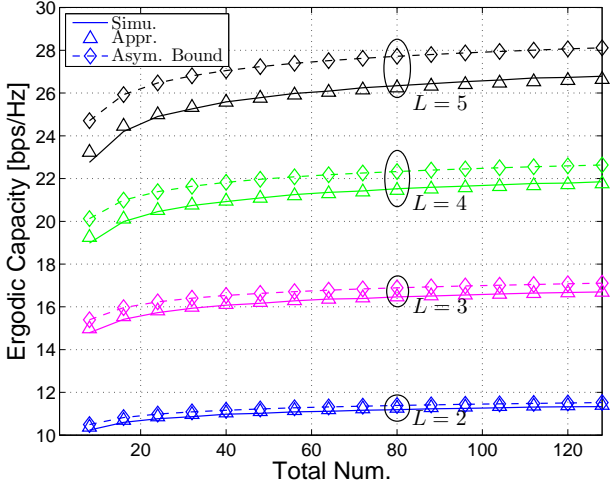} 
\caption{Simulated (solid line), approximated ($\triangle$) and Upper-bounded ergodic capacity ($\Diamond$) versus the total number of acquired CSI when $L$ increases from 2 to 5. $N_{\rm{r}}=128$, $N_{\rm{t}}=8$ and $\tilde{\rho}$ is 5dB.}
\label{figure4}
\end{figure}

\begin{figure*}[!t] 
    \centering
    \subfigure[$\tilde{\rho}=-10~\text{dB}$]
    {
        \includegraphics[width=0.33\textwidth]{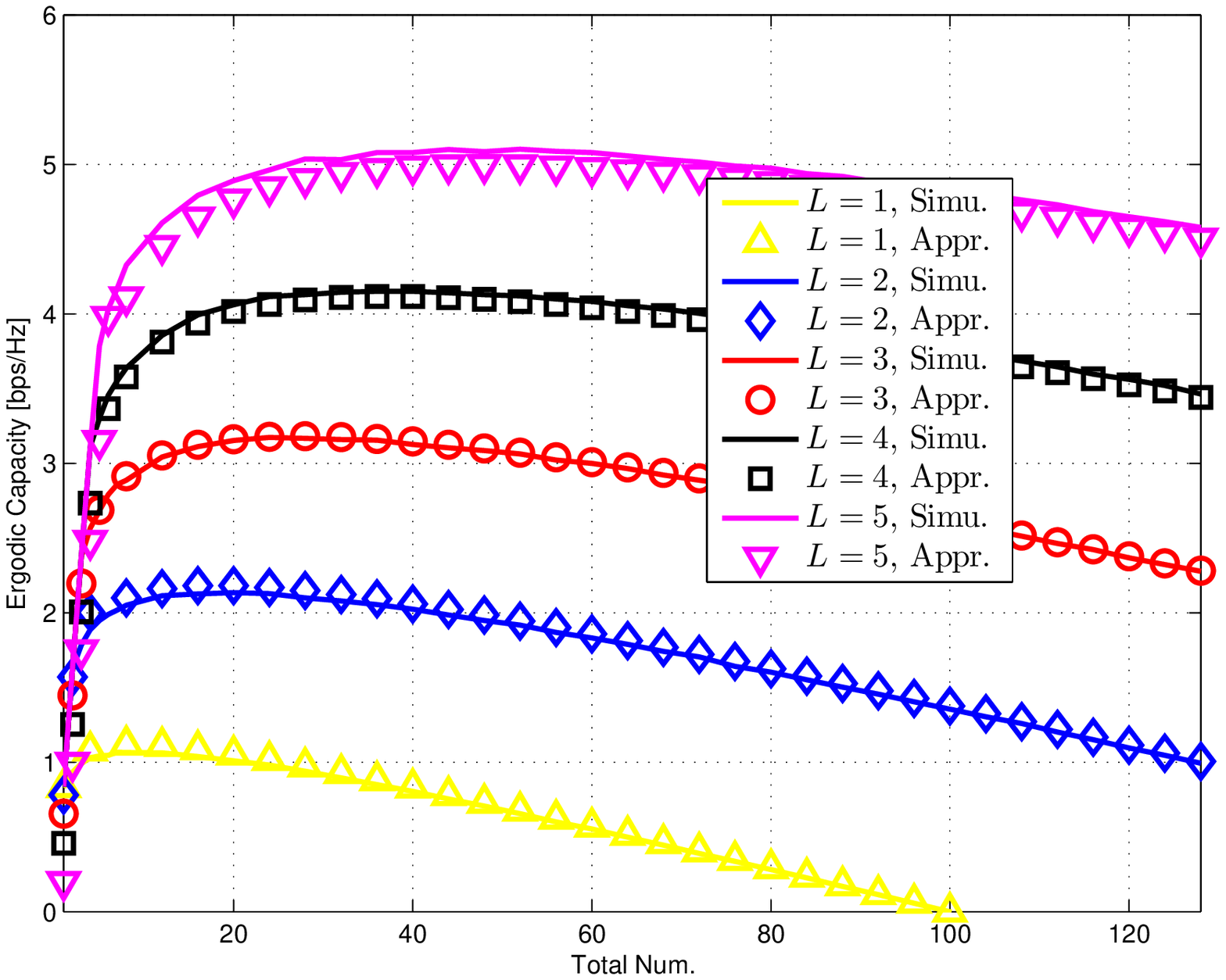}
	   \label{fig5a}	   
    } 
\vspace{-5pt}
\hspace{-15pt}
   \subfigure[$\tilde{\rho}=0~\text{dB}$]
    {
        \includegraphics[width=0.33\textwidth]{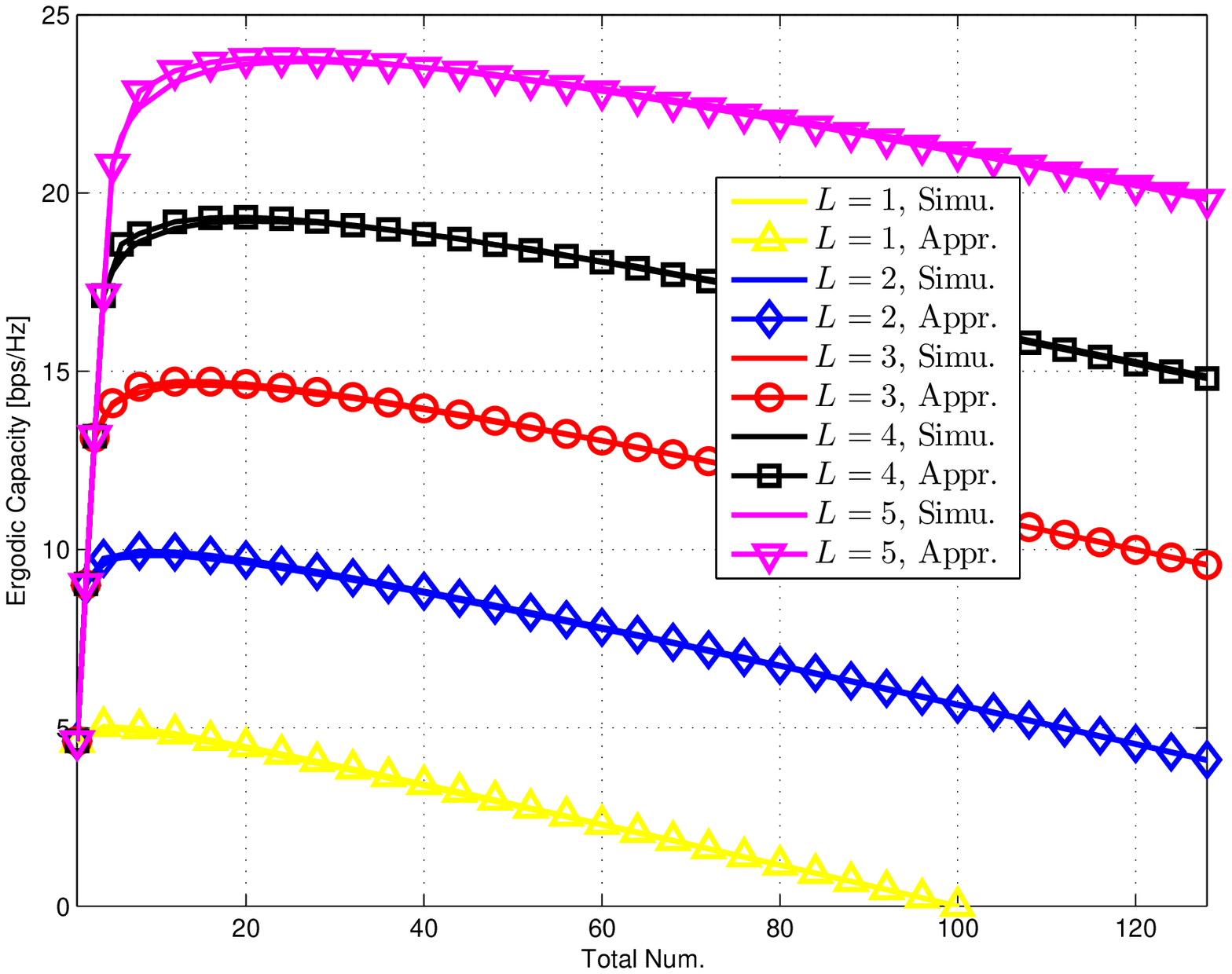}
	   \label{fig5b}	   
    } 
\vspace{-5pt}
\hspace{-15pt}
    \subfigure[$\tilde{\rho}=10~\text{dB}$]
    {
        \includegraphics[width=0.33\textwidth]{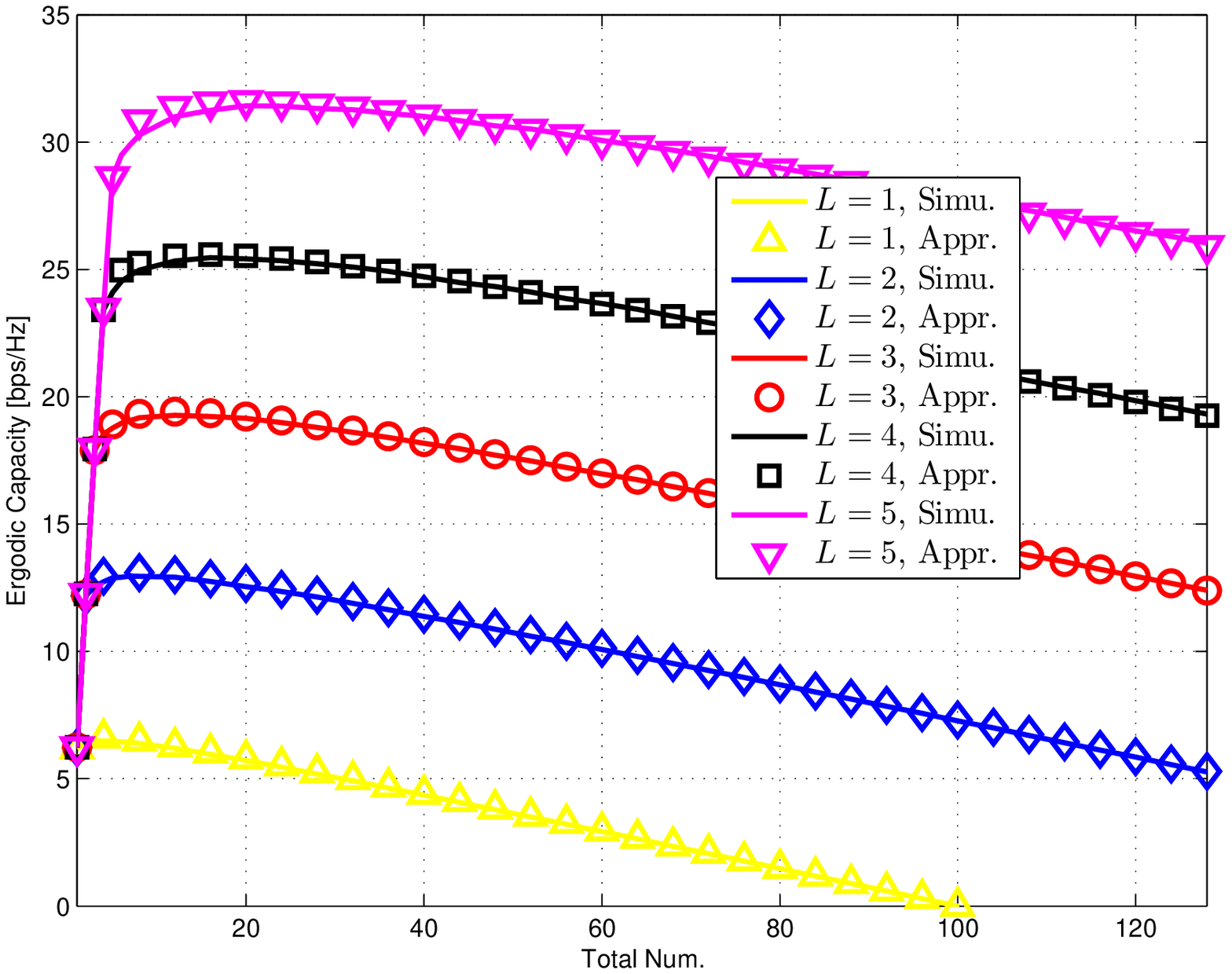}
	   \label{fig5c}	   
    } 
\\
    \caption{Efficient ergodic capacity versus the total number of the acquired CSI. $\tilde{\rho}$ increase from -10 dB to 10 dB when $N_{\rm{r}}=128$ and $N_{\rm{t}}=8$.}
    \label{figure5}
	\vspace{0.2in}
\end{figure*}

{\figurename} \ref{figure4} shows the exact, approximated and upper-bounded ergodic channel capacity. The horizontal axis represents the total number (Total Num.) of acquired CSI, i.e., $\Upsilon$. Actually,  $\Upsilon<N_{\rm{r}}$ is the case when only $\Upsilon$ out of $N_{\rm{r}}$ rows of the channel matrix is available at the receiver, which means only partial CSI is used during the antenna selection. As a result, the final $L$ activated antennas are selected from the full set constructed by the $\Upsilon$ antennas. In this figure, the approximated channel capacity is calculated by \eqref{equ20} and the upper-bounded result is the asymptotic approximated upper bound. The approximated and simulated ergodic capacity are nearly equivalent for different $L$ which means the empirical formula \eqref{equ18} is reasonable. In addition, the ergodic capacity rises slowly with the increment of acquired CSI, which further supports the findings observed in {\figurename} \ref{figure3}.

Next, we will turn to the efficient capacity defined before. $\Upsilon$ is the amount of acquired CSI before the antenna selection, then the time consumed for CSI acquisition is $\frac{\Upsilon}{L}t_{\rm{tr}}$, thus the efficient ergodic capacity is given by
\begin{equation}
\begin{split}
{\hat{C}_{\rm{s,eff}}}&={\hat{C}_{\rm{s}}}\left(1-\frac{\Upsilon}{L}\eta\right),
\end{split}
\end{equation}
where $\eta=\frac{t_{\rm{tr}}}{t_{\rm{coh}}}$ and ${\hat{C}_{\rm{s}}}$ is ergodic capacity for AS-MIMO system when only partial CSI is used. It is apparent ${\hat{C}_{\rm{s}}}$ will increases with $\Upsilon$, whereas $\left(1-\frac{\Upsilon}{L}\eta\right)$ is on the contrary. Therefore, there seems to be an optimal $\Upsilon$ which may maximize the efficient capacity. {\figurename} \ref{figure5} presents the exact and approximated efficient capacity versus $\Upsilon$ when the total number of receive antenna is $N_{\rm{r}}=128$ and $\eta=0.01$. In each sub-figure, $\Upsilon$ increases from 1 to 128 and $\tilde{\rho}$ ranges from -10 dB to 10 dB. The most interesting aspect of this graph is that there exists an optimal $\Upsilon$ for different $L$ and normalized SNR. When $\tilde{\rho}$ is fixed, the optimal number of acquired CSI increases with $L$. In addition, when $L$ is fixed, the optimal value of $\Upsilon$ decreases when $\tilde{\rho}$ rises up. In contrast to earlier findings in {\figurename} \ref{figure4}, it is evident that CSI acquisition has apparent influence on the performance of antenna selection. If the receiver is equipped with a much larger antenna array, CSI acquisition would place an enormous burden on system performance, lowering the spectral efficiency and 
energy efficiency of the massive MIMO system. 

By now, assisted by the asymptotic approximated upper bound, this section has reviewed two key aspects of antenna selection. These are:
\begin{itemize}
\item 
The ergodic capacity of AS-MIMO systems increases slowly with the total number of acquired CSI.
\item
CSI acquisition has great influence on the performance of AS-MIMO systems, lowering its efficiency.
\end{itemize}
The chapter that follows moves on to explore the inherent characteristics of AS-MIMO systems and utilize it to design a much more efficient selection algorithm.

\section{Antenna Selection with Partial CSI}
\label{sec5} 
\subsection{$L\leq{N_{\rm{t}}}$}
\begin{figure}[!h] 
    \centering
    \subfigure[$N_{\rm{r}}=128$, $N_{\rm{t}}=4$, $L=4$]
    {
        \includegraphics[width=0.45\textwidth]{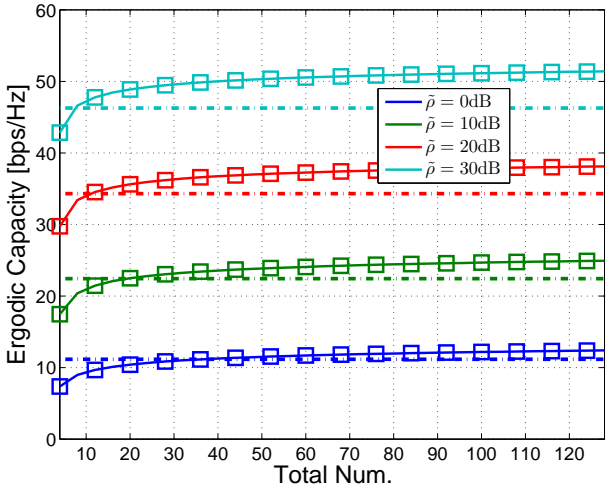}
	   \label{fig6a}	   
    } 
\\
    \subfigure[$N_{\rm{r}}=128$, $N_{\rm{t}}=8$, $L=4$]
    {
        \includegraphics[width=0.45\textwidth]{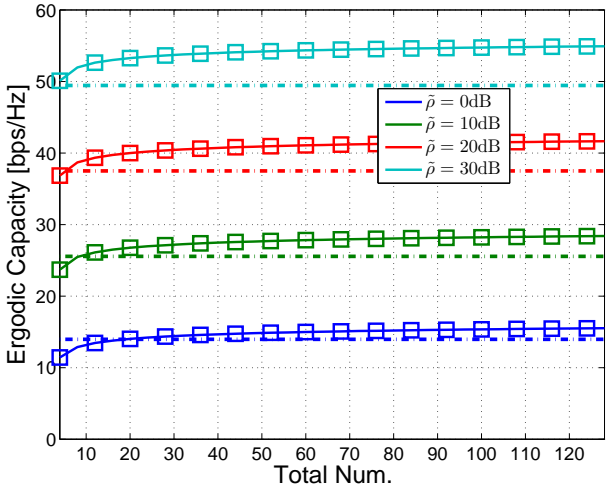}
      \label{fig6b}	        
    }
\\
    \caption{Ergodic capacity ($\Box$) versus the total number (Total Num.) of acquired CSI at different normalized SNR. $N_{\rm{t}}$ takes 4 and 8 when $N_{\rm{r}}=128$ and $L=4$. $\tilde{\rho}$ increases from 0 dB to 30 dB which are denoted by different colors. The dot-dash lines represent the 0.9-Level of the largest ergodic capacity by exhaustive search when full CSI is utilized.}
    \label{figure6}
\end{figure}

\begin{figure}[!t] 
\setlength{\abovecaptionskip}{0pt} 
\centering 
\includegraphics[width=0.45\textwidth]{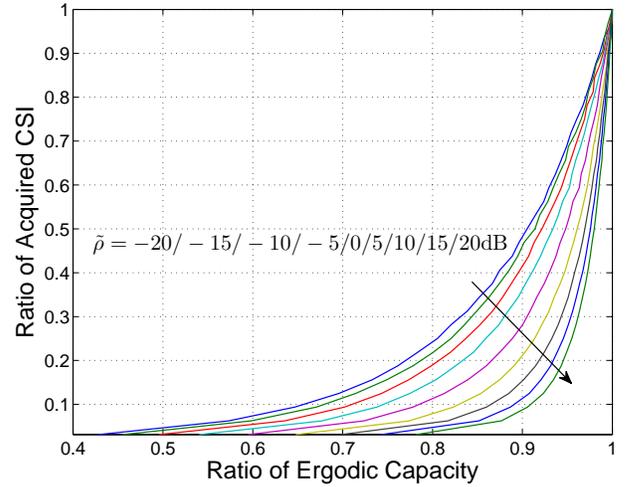} 
\caption{The relationship between acquired CSI and achievable ergodic capacity when $N_{\rm{r}}=128$, $N_{\rm{t}}=4$ and $L=4$. $\tilde{\rho}$ increases from -20 dB to 20 dB.}
\label{figure7}
\end{figure}

As was mentioned in the previous chapter, the ergodic capacity grows at a sluggish pace when $\Upsilon$ rises up. This phenomenon can be observed both in {\figurename} \ref{figure3} and {\figurename} \ref{figure4}. To express this characteristic more clearly, {\figurename} \ref{figure6} compares the ergodic capacity (obtained by Monte-Carlo simulation) with partial CSI (denoted by $\Box$) and 0.9-Level of the achievable rate with full CSI (denoted by dot-dash line) in different set-up. As can be seen from {\figurename} \ref{fig6a} and {\figurename} \ref{fig6b}, $\Upsilon=40$ is enough in terms of reaching the 0.9-Level of the largest achievable capacity. Let $\Upsilon^*$ denote the total number of acquired CSI when the 0.9-Level is attained. As can be seen from these two sub-figures, $\Upsilon^*$ decreases as $\tilde{\rho}$ grows, which means less CSI is required in order to achieve the same channel capacity when the channel quality is better. Furthermore, {\figurename} \ref{figure7} shows the relationship between the acquired CSI and achievable channel capacity when $\tilde{\rho}$ ranges from -20 dB to 20 dB. The vertical axis represents the ratio of achievable channel capacity with partial CSI to the largest capacity with full CSI i.e., $r_1=\hat{C}_{\rm{s}}/{C}_{\rm{s}}$ and the horizontal axis denotes the ratio of acquired CSI to the total number of receive antennas i.e., $r_2={\Upsilon}/{N_{\rm{r}}}$. As can be seen from this graph, the relation ship between the achievable transmission rate and acquired CSI follows the Pareto principle approximately, i.e., 80\% of the effects come from 20\% of the causes. In our case, 80\% of the largest channel capacity can be reached via only 20\% of the CSI once $\tilde{\rho}\geq-10~\rm{dB}$ .

In fact, utilization of full CSI will cause two challenges for antenna selection. These are:
\begin{itemize}
\item 
Acquisition of full CSI will consume too much resources, lowering the spectral and energy efficiency of the whole system.
\item
Full CSI means that $L$ antennas will be selected from a full set with large size. Since there has been no efficient optimal selection algorithm proposed, this will cause prohibitive complexity.  
\end{itemize}
Moreover, as previously stated, the relationship between the acquired CSI and achievable channel capacity follows Pareto principle. On the basis of these factors, an adaptive selection algorithm can be designed which only acquires partial CSI but can obtain a remarkable performance. 

The basic idea of this algorithm is to gradually increase the number of acquired CSI which is used for antenna selection until the achievable channel capacity exceeds a predefined value. Let ${\mathcal{S}}_{\rm{n}}$ represent the full set of antenna indexes, the corresponding rows of which in the channel matrix have been acquired until the $n$-th step. In addition, let ${\mathcal{I}}_{\rm{n}}$ denotes the set of acquired CSI i.e., the row indexes of the channel matrix at the $n$-th step. Therefore, it is apparent that ${\mathcal{S}}_{{\rm{n}}+1}={\mathcal{S}}_{{\rm{n}}}\cup{\mathcal{I}}_{\rm{n}+1}$ holds. After ${\mathcal{I}}_{\rm{n}}$ is obtained, antenna selection is applied in ${\mathcal{S}}_{{\rm{n}}}$ to find the largest achievable channel capacity which is termed as $C_{\rm{n}}$. Next, ${\mathcal{I}}_{{\rm{n}}+1}$ is acquired and the antenna selection among the full set ${\mathcal{S}}_{{\rm{n}}+1}$ will go on. This `CSI Acquisition + Antenna Selection' procedure will not stop until the achievable result $C_{\rm{n}}$ reaches the predefined value. The whole algorithm is summarized in Alg. \ref{alg1}.

\begin{algorithm}[!t]
\caption{Adaptive Antenna Selection when $N_{\rm{t}}\geq{L}$}
\label{alg1}
\begin{algorithmic}
\Require ~~\\
The predefined value of the target channel capacity, $C_{\rm{final}}$;
\Ensure ~~\\
The total number of acquired CSI, $\Upsilon_{\rm{A}}$; 
\\The optimal antenna subset, ${\mathcal{T}}_{\rm{op}}$; 
\\The achievable channel capacity, $C_{\rm{ac}}$; 
\end{algorithmic}
\begin{algorithmic}[1]
\State ${\mathcal{S}}_{\rm{0}}=\phi$, ${\mathcal{I}}_{\rm{0}}=\phi$, ${{C}}_{\rm{0}}=0$, ${\rm{n}}=0$, ${\mathcal{T}}_{\rm{0}}=\phi$ 
\While {$C_{\rm{n}}<{C_{\rm{final}}}$}
\State ${\rm{n}} \leftarrow {\rm{n}}+1$
\State ${\mathcal{S}}_{{\rm{n}}+1}={\mathcal{S}}_{{\rm{n}}}\cup{\mathcal{I}}_{\rm{n}+1}$
\State Select the optimal antenna subset ${\mathcal{T}}_{{\rm{n}}+1}$ from ${\mathcal{S}}_{{\rm{n}}+1}$ 
\State Calculate $C_{{\rm{n}}+1}$ based on ${\mathcal{T}}_{{\rm{n}}+1}$ 
\EndWhile
\State $\Upsilon_{\rm{A}}=|\mathcal{{S}}_{{\rm{n}}}|$  
\State ${\mathcal{T}}_{\rm{op}}={\mathcal{T}}_{\rm{n}}$
\State $C_{\rm{ac}}=C_{{\rm{n}}}$\\
\Return $\Upsilon_{\rm{A}}$, ${\mathcal{T}}_{\rm{op}}$, $C_{\rm{ac}}$
\end{algorithmic}
\end{algorithm}

As stated before, CSI acquisition is done by RF chains, thus the number of acquired CSI at the $n$-th step can not exceed the total number of RF chains i.e., $|{\mathcal{I}}_{\rm{n}}|\leq L$. In Alg. \ref{alg1}, the antenna selection in the 5th line can be based on exhaustive search (ES) or branch-and-bound search (BAB), since both these two algorithms can achieve the optimal solution. Attainment of the predefined value $C_{\rm{final}}$ marks the end Alg. \ref{alg1}, and this value depends on the actual demand. One method to calculate this value is using \eqref{equ20}. For example, $C_{\rm{final}}$ can be set to $0.9\left({\bar{C}}_{\rm{asy}}+\tilde{\Xi}\right)$ if the goal is to reach the 0.9-Level of the largest achievable transmission rate. Two key challenges followed by full CSI acquisition in AS-MIMO systems have been investigated before, and these two challenges can be both addressed by the new proposed adaptive algorithm. More specifically, only partial CSI is required, which can alleviate the requirement of resources, such as time and energy, in CSI acquisition. On the other hand, the size of the candidate antenna set is much smaller, which may result in a low computation complexity.  

\begin{figure*}[!t] 
    \centering
    \subfigure[$N_{\rm{r}}=64$, $N_{\rm{t}}=8$, $L=5$]
    {
        \includegraphics[width=0.45\textwidth]{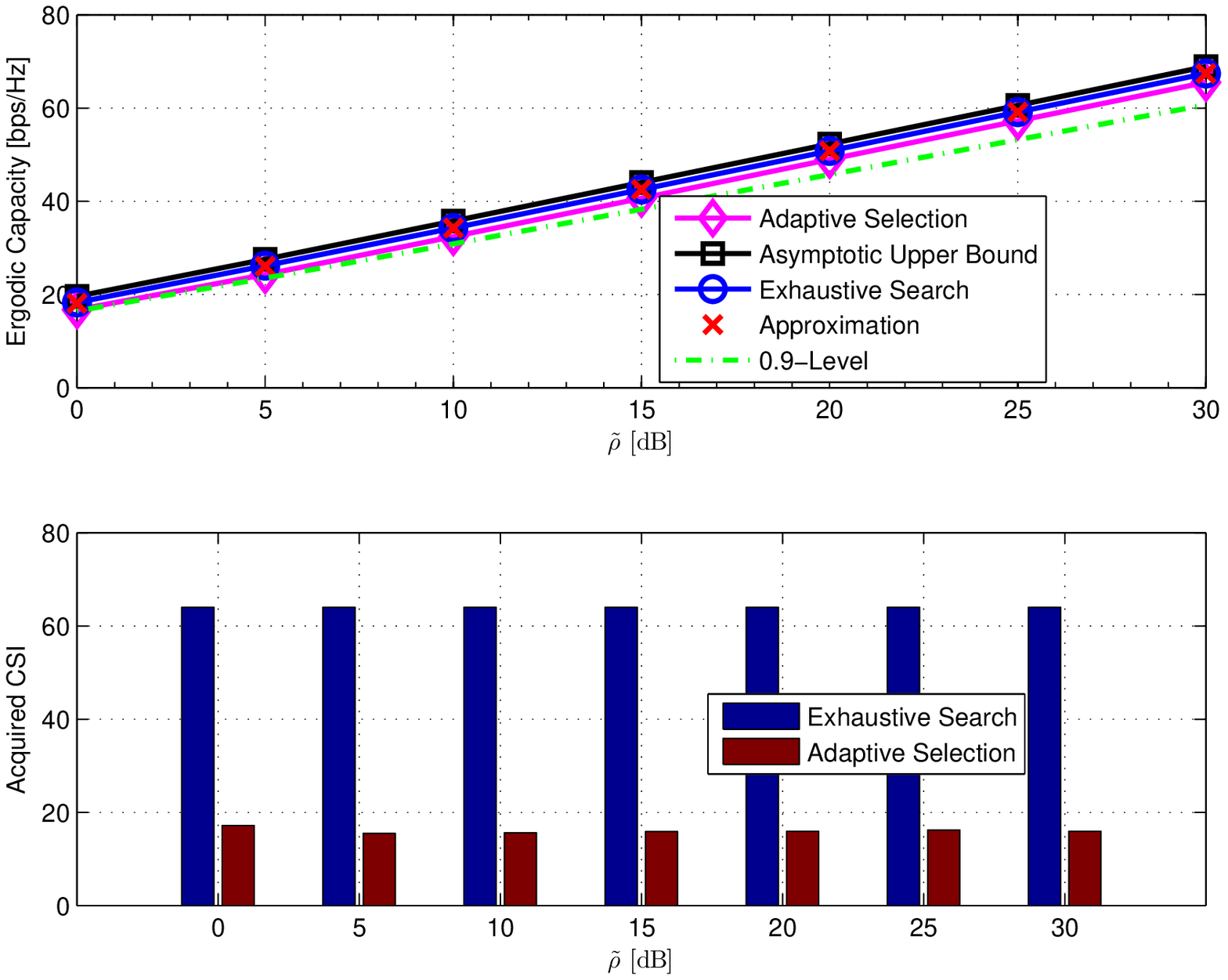}
	   \label{fig8a}	   
    } 
\vspace{-5pt}
\hspace{-15pt}
   \subfigure[$N_{\rm{r}}=100$, $N_{\rm{t}}=7$, $L=5$]
    {
        \includegraphics[width=0.45\textwidth]{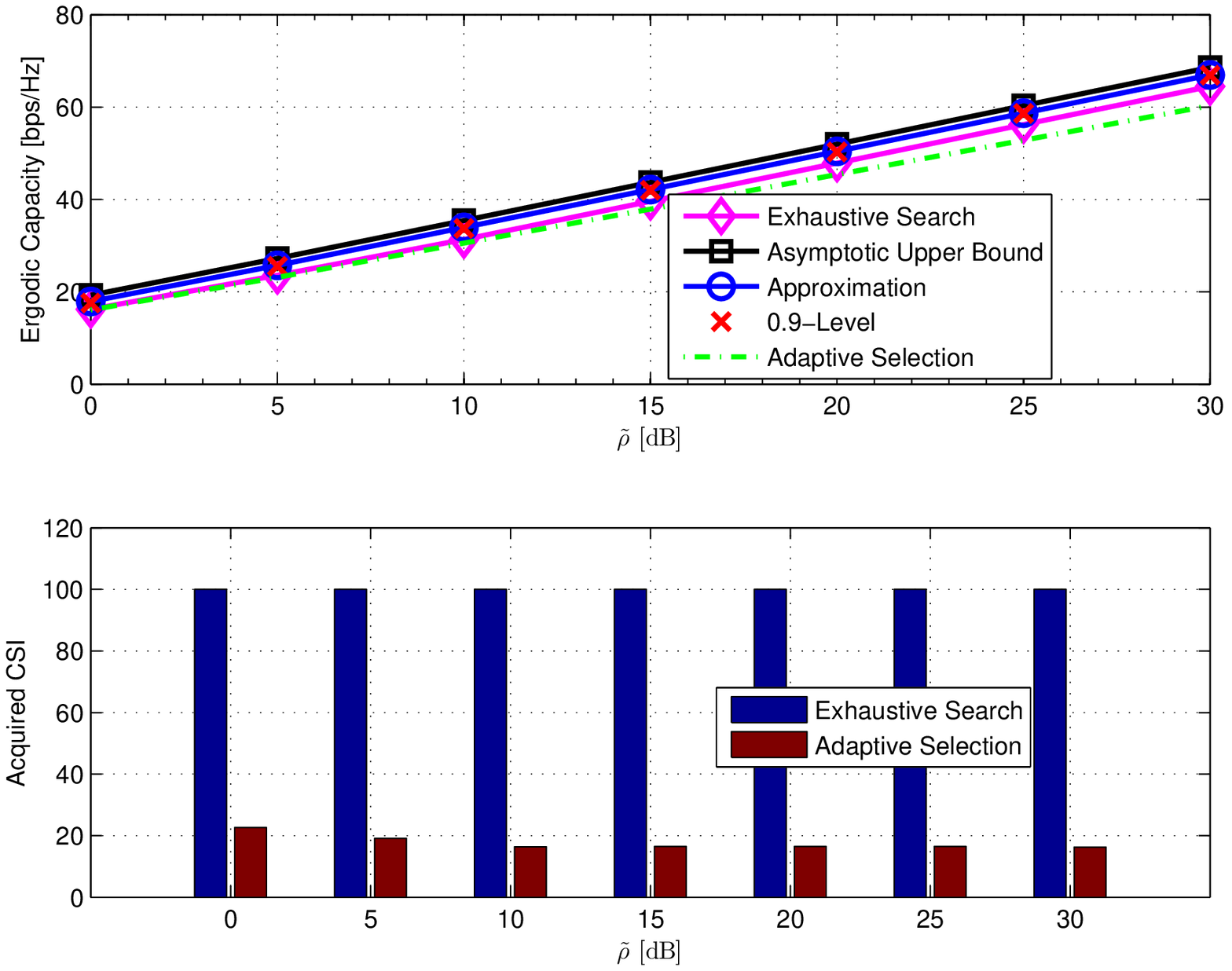}
	   \label{fig8b}	   
    } \\
    \subfigure[$N_{\rm{r}}=128$, $N_{\rm{t}}=4$, $L=4$]
    {
        \includegraphics[width=0.45\textwidth]{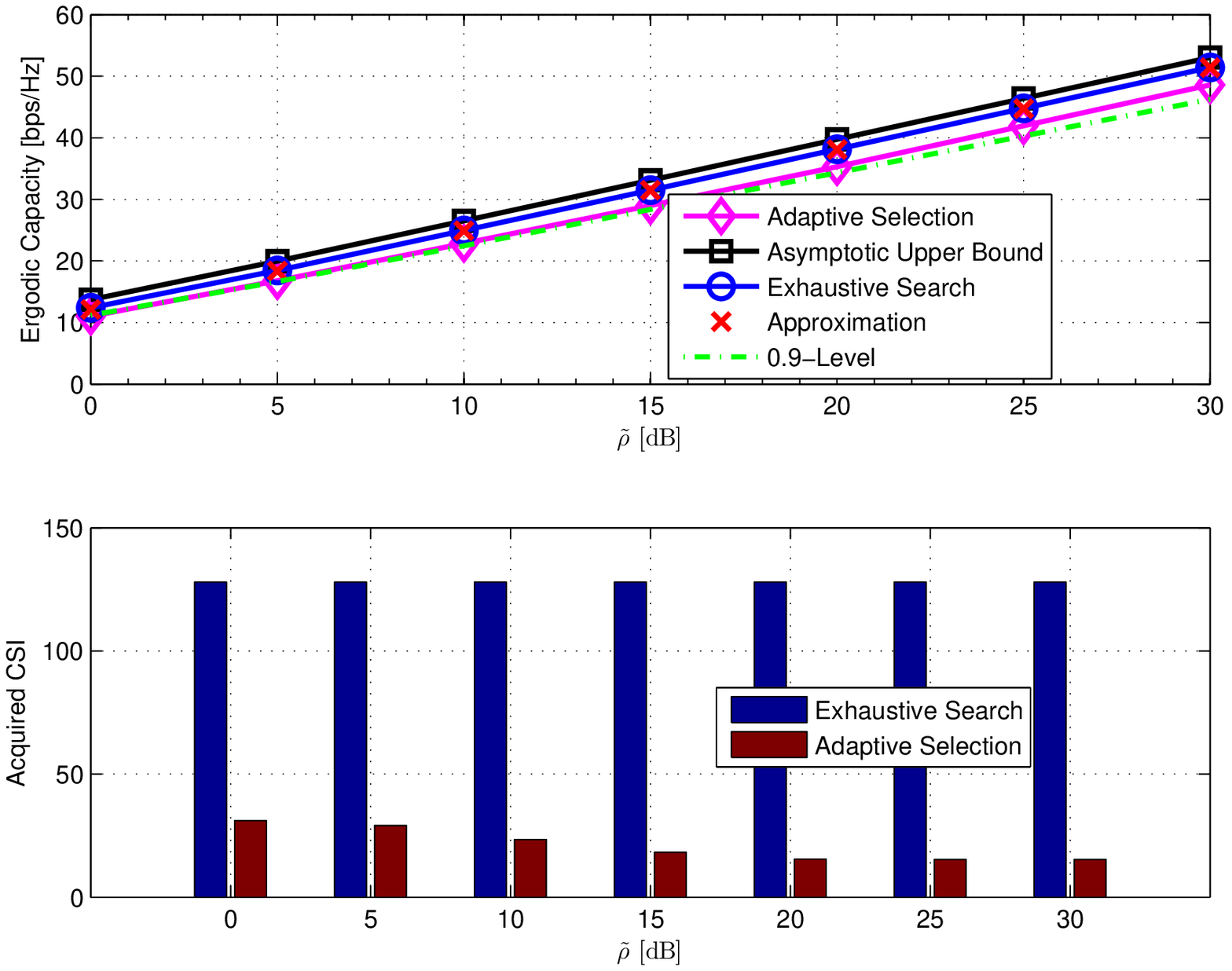}
	   \label{fig8c}	   
    } 
\vspace{-5pt}
\hspace{-15pt}    
    \subfigure[$N_{\rm{r}}=128$, $N_{\rm{t}}=8$, $L=4$]
    {
        \includegraphics[width=0.45\textwidth]{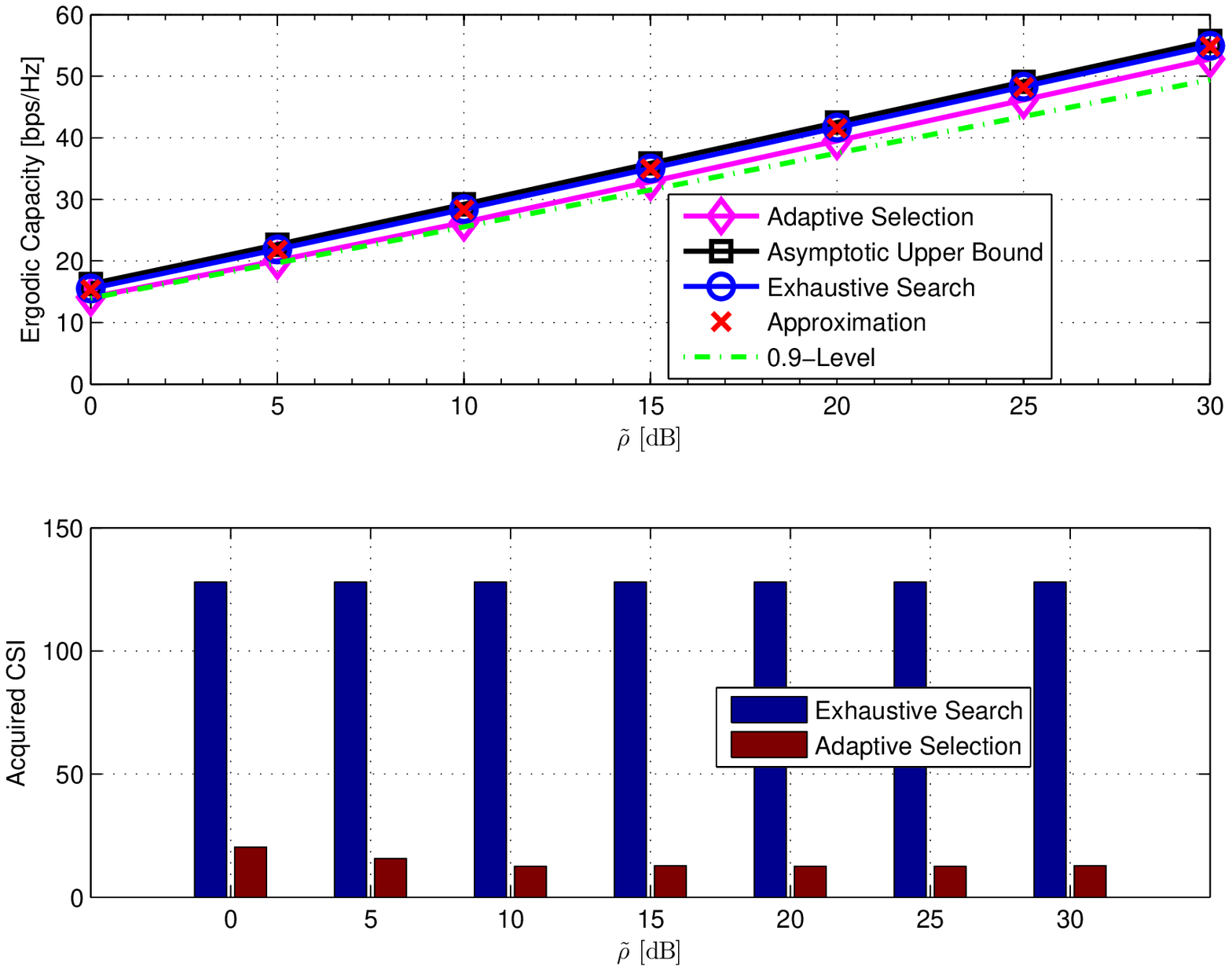}
      \label{fig8d}	        
    }
\\
    \caption{Asymptotic upper bound ($\Box$), ES-based ergodic capacity($\circ$), approximated ES-based ergodic values ($\times$), adaptive selection based ergodic capacity ($\triangleleft$) and 0.9-Level ($\cdot-\cdot$) of ES-based ergodic capacity versus $\tilde{\rho}$ for different antenna deployment styles are illustrated at the top of each sub-figure. The mean amount of acquired CSI verus $\tilde{\rho}$ during the whole algorithm for both exhaustive search and adaptive search are plotted at the bottom of each sub-figure}
    \label{figure8}
    \vspace{-0.1in}
\end{figure*} 

{\figurename} \ref{figure8} presents the ergodic capacity and the mean number of acquired CSI of the proposed adaptive algorithm for different antenna deployment styles. In each sub-figure, the top graph compares the ergodic value of the adaptive search and exhaustive search and the graph below shows the number of acquired CSI versus $\tilde{\rho}$. In each scenarios, the predefined goal is set to be the 0.9-Level of the largest capacity obtained by ES. As can be seen from this graph, the adaptive search can arrive at the preset 0.9-Level and the cardinality of ${\mathcal{I}}_{\rm{n}}$ is $|{\mathcal{I}}_{\rm{n}}|=L$. In addition, the number of acquired CSI is much smaller than that of the exhaustive search. Furthermore, there is a decreasing trend of this number as $\tilde{\rho}$ grows, which further supports the observed Pareto principle. 

It should be noticed that ergodic capacity is used as the metric of comparison in {\figurename} \ref{figure8}. Next, let us turn to discuss the efficient capacity defined in the previous section. As shown in {\figurename} \ref{fig8b}, the total number of acquired CSI is no larger than 20 once $\tilde{\rho}>20\rm{dB}$ when $N_{\rm{r}}=100$ and $N_{\rm{t}}=7$. The efficient capacity for the adaptive search and exhaustive search can be calculated by $C_{\rm{ac}}\left(1-\frac{\Upsilon_{\rm{A}}}{L}\eta\right)$ and $C_{\rm{final}}\left(1-\frac{100}{L}\eta\right)$, respectively. On the basis of the simulation results, the relationship $C_{\rm{ac}}>0.9C_{\rm{final}}$ and $\Upsilon_{\rm{A}}<20$ hold. Suppose that $\eta$ is set to be 0.01, then the following inequality is satisfied:
\begin{equation}
\begin{split}
C_{\rm{ac}}\left(1-\frac{\Upsilon_{\rm{A}}}{L}\eta\right)&>C_{\rm{final}}\times0.9\left(1-\frac{20}{5}\times0.01\right)\\&=0.864C_{\rm{final}}>0.8C_{\rm{final}}\\&=C_{\rm{final}}\left(1-\frac{100}{L}\eta\right)
\end{split}
\end{equation}
Therefore, the efficient ergodic capacity of the adaptive search is larger than that of the exhaustive search. Actually, this is only a special example. Once $\eta$ equals to another value, the efficient capacity of the ES may exceed that of the adaptive search. Nevertheless, this problem can be simply solved by setting $C_{\rm{final}}$ to a new value. 

\begin{figure}[!t] 
\setlength{\abovecaptionskip}{0pt} 
\centering 
\includegraphics[width=0.45\textwidth]{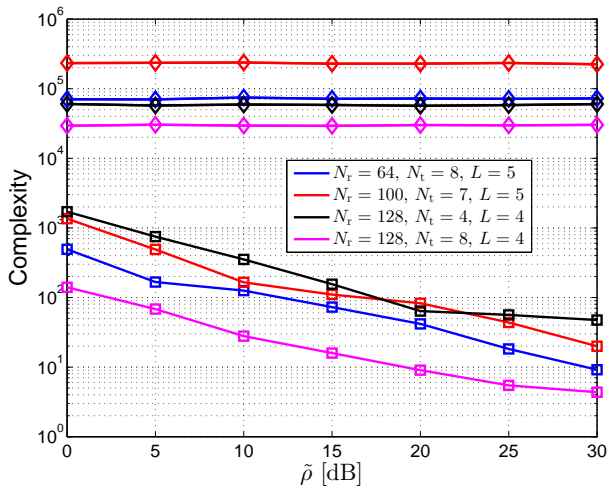} 
\caption{Complexity versus $\tilde{\rho}$ for both branch-and-bound based search ($\Diamond$) and adaptive search ($\Box$) in different antenna deployment styles.}
\label{figure9}
\end{figure} 

{\figurename} \ref{figure9} compares the computation complexity of the antenna selection algorithm with full CSI (by branch-and-bound search) and partial CSI (by adaptive search). The antenna selection algorithm used in the adaptive search are based on the branch-and-bound method and the final goal of the adaptive search is set to be $0.9\left({\bar{C}}_{\rm{asy}}+\tilde{\Xi}\right)$. The complexity is defined as the total number of visited nodes during the tree search of the branch-and-bound search. It can be seen from {\figurename} \ref{figure9} that the adaptive search posses a much lower complexity than the branch-bound-search search which requires full CSI. Moreover, the complexity decreases with the increment of $\tilde{\rho}$, which is due to the decreasing of the total number of acquired CSI. 

This section has demonstrated the situation when $N_{\rm{t}}\geq{L}$. It is now necessary to explain the course of $N_{\rm{t}}<{L}$.

\subsection{$L>{N_{\rm{t}}}$}
\begin{figure}[!t] 
\setlength{\abovecaptionskip}{0pt} 
\centering 
\includegraphics[width=0.45\textwidth]{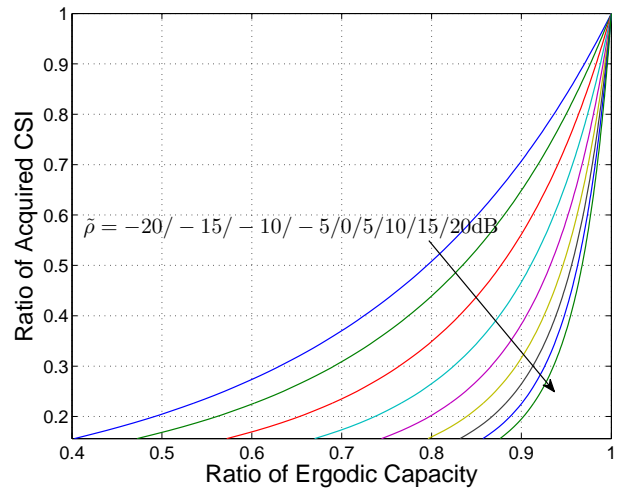} 
\caption{The relationship between acquired CSI and reachable ergodic capacity when $N_{\rm{r}}=128$, $N_{\rm{t}}=8$ and $L=20$. $\tilde{\rho}$ increases from -20 dB to 20 dB.}
\label{figure10}
\end{figure}

Since it is hard to find the optimal antenna subset when $L$ is large, the derived asymptotic upper bound (calculated by \eqref{mx}) can be used to explore the relationship between the achievable transmission rate and the amount of acquired CSI, which is illustrated in {\figurename} \ref{figure10}. As shown in this graph, the relationship between the channel capacity and acquired CSI in AS-MIMO systems still follows the similar law as Pareto principle. On the basis of this characteristic, an adaptive algorithm can be also designed when $L>{N_{\rm{t}}}$. Nevertheless, there are two problems, these are:
\begin{itemize}
\item
There is no precise approximation for the exact ergodic capacity when $L>{N_{\rm{t}}}$, thus the predefined goal for the adaptive search is hard to obtain.
\item
Both the Exhaustive search and branch-and-bound search are of high-complexity when $L$ is large, thus optimal selection algorithm can not be used in the adaptive search procedure. 
\end{itemize}  
In regard to these two problems, two relaxed strategy can be utilized. As stated before, greedy search can achieve near optimal performance with much lower complexity than ES and BAB. Therefore, greedy search can serve as the selection algorithm in the whole adaptive procedure. Moreover, the asymptotic upper bound is tight when $L$ is much larger than $N_{\rm{t}}$, thus it makes sense to use \eqref{mx} to predetermine the final goal of the whole algorithm. 

The whole algorithm for the scenario $L>{N_{\rm{t}}}$ is summarized in Alg. \ref{alg2} which is similar as Alg. \ref{alg1}. It should be noticed that the antenna selection algorithm in the 5th line is fixed to be greedy search in this new situation. 

\begin{algorithm}[!t]
\caption{Adaptive Antenna Selection when $N_{\rm{t}}<{L}$}
\label{alg2}
\begin{algorithmic}
\Require ~~\\
The preset value of the objective channel capacity, $C_{\rm{final}}$;
\Ensure ~~\\
The total number of acquired CSI, $\Upsilon_{\rm{A}}$; 
\\The optimal antenna subset, ${\mathcal{T}}_{\rm{op}}$; 
\\The achievable channel capacity, $C_{\rm{ac}}$; 
\end{algorithmic}
\begin{algorithmic}[1]
\State ${\mathcal{S}}_{\rm{0}}=\phi$, ${\mathcal{I}}_{\rm{0}}=\phi$, ${{C}}_{\rm{0}}=0$, ${\rm{n}}=0$, ${\mathcal{T}}_{\rm{0}}=\phi$ 
\While {$C_{\rm{n}}<{C_{\rm{final}}}$}
\State ${\rm{n}} \leftarrow {\rm{n}}+1$
\State Acquire CSI to obtain ${\mathcal{I}}_{\rm{n}+1}$ (${\mathcal{S}}_{\rm{n}}\cap{\mathcal{I}}_{\rm{n}+1}=\phi$)
\State ${\mathcal{S}}_{{\rm{n}}+1}={\mathcal{S}}_{{\rm{n}}}\cup{\mathcal{I}}_{\rm{n}+1}$
\State Select ${\mathcal{T}}_{{\rm{n}}+1}$ from ${\mathcal{S}}_{{\rm{n}}+1}$ by greedy search
\State Calculate $C_{{\rm{n}}+1}$ 
\EndWhile
\State $\Upsilon_{\rm{A}}=|\mathcal{{S}}_{{\rm{n}}}|$  
\State ${\mathcal{T}}_{\rm{op}}={\mathcal{T}}_{\rm{n}}$
\State $C_{\rm{ac}}=C_{{\rm{n}}}$\\
\Return $\Upsilon_{\rm{A}}$, ${\mathcal{T}}_{\rm{op}}$, $C_{\rm{ac}}$
\end{algorithmic}
\end{algorithm}

\begin{figure*}[!t] 
    \centering
    \subfigure[$N_{\rm{r}}=64$, $N_{\rm{t}}=8$, $L=19$]
    {
        \includegraphics[width=0.45\textwidth]{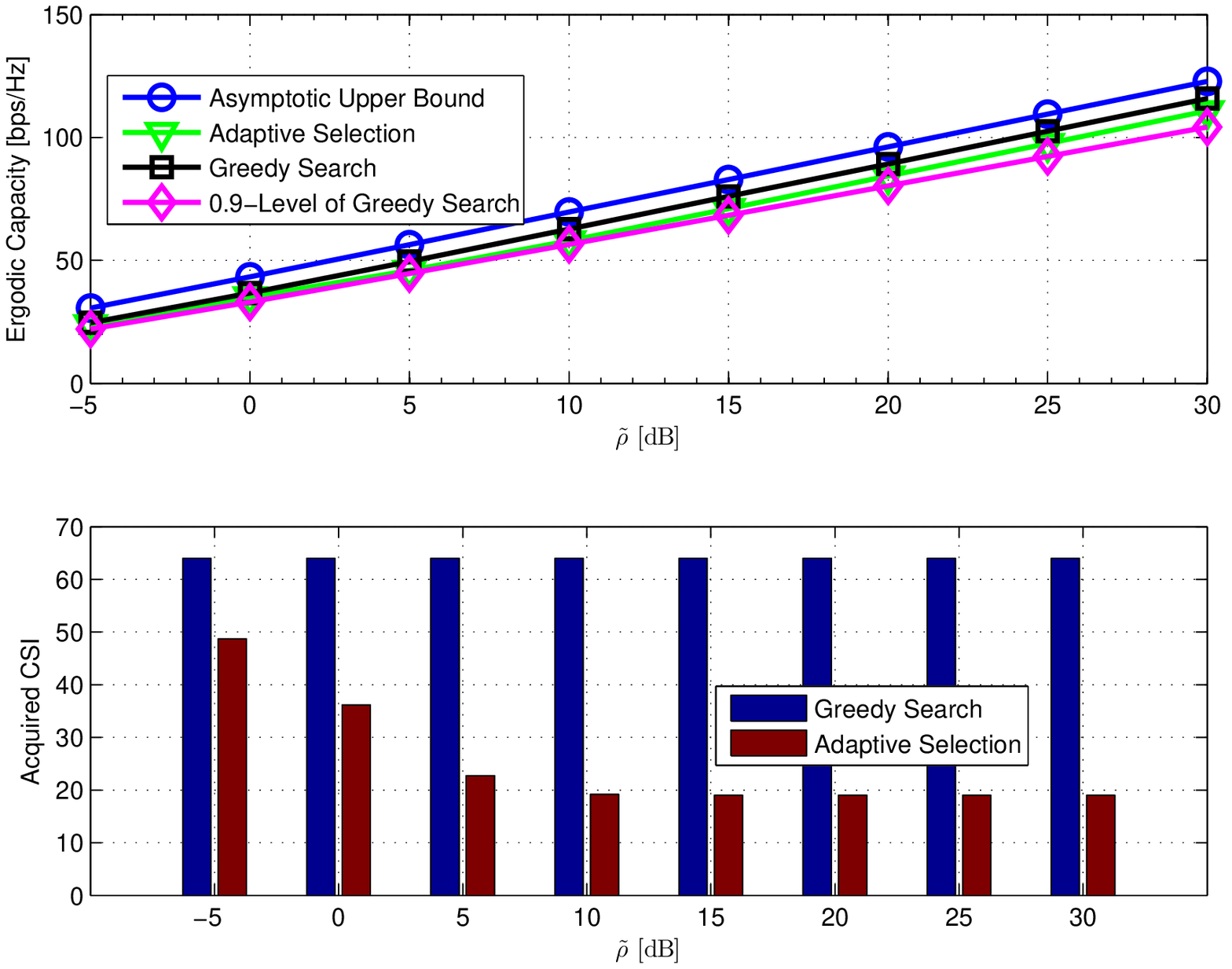}
	   \label{fig11a}	   
    } 
\vspace{-5pt}
\hspace{-15pt}
   \subfigure[$N_{\rm{r}}=100$, $N_{\rm{t}}=7$, $L=16$]
    {
        \includegraphics[width=0.45\textwidth]{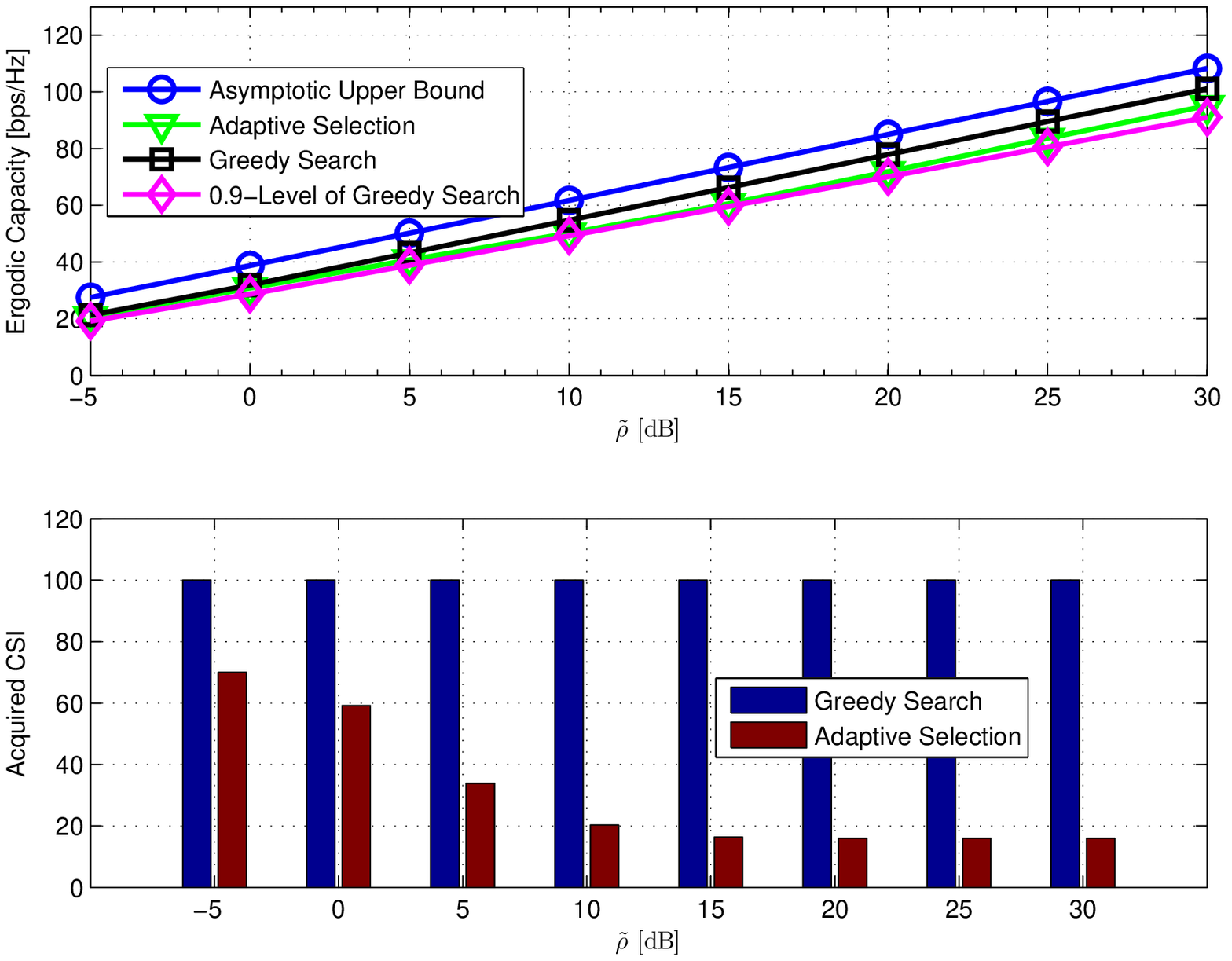}
	   \label{fig11b}	   
    } \\
    \subfigure[$N_{\rm{r}}=128$, $N_{\rm{t}}=4$, $L=16$]
    {
        \includegraphics[width=0.45\textwidth]{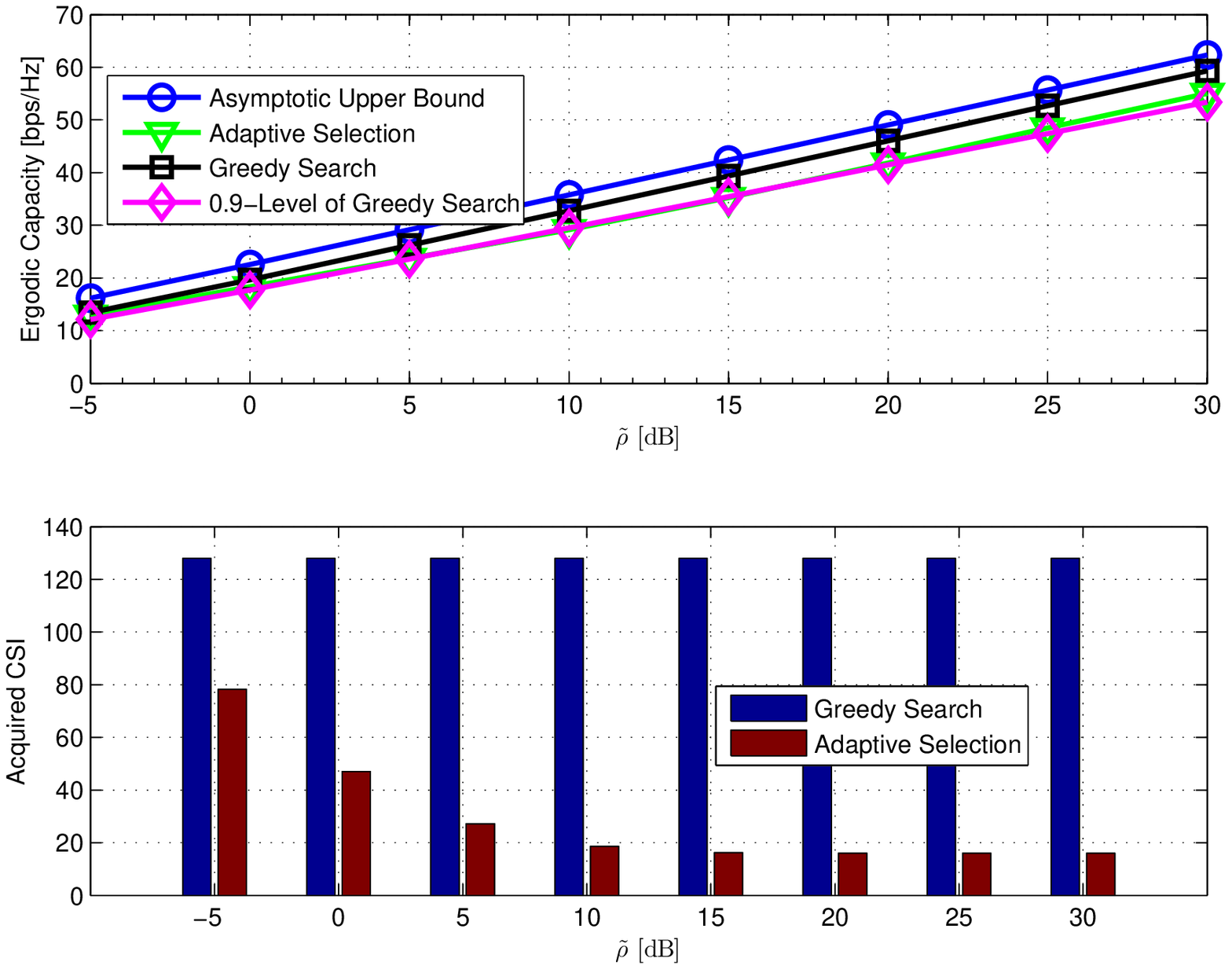}
	   \label{fig11c}	   
    } 
\vspace{-5pt}
\hspace{-15pt}    
    \subfigure[$N_{\rm{r}}=128$, $N_{\rm{t}}=8$, $L=20$]
    {
        \includegraphics[width=0.45\textwidth]{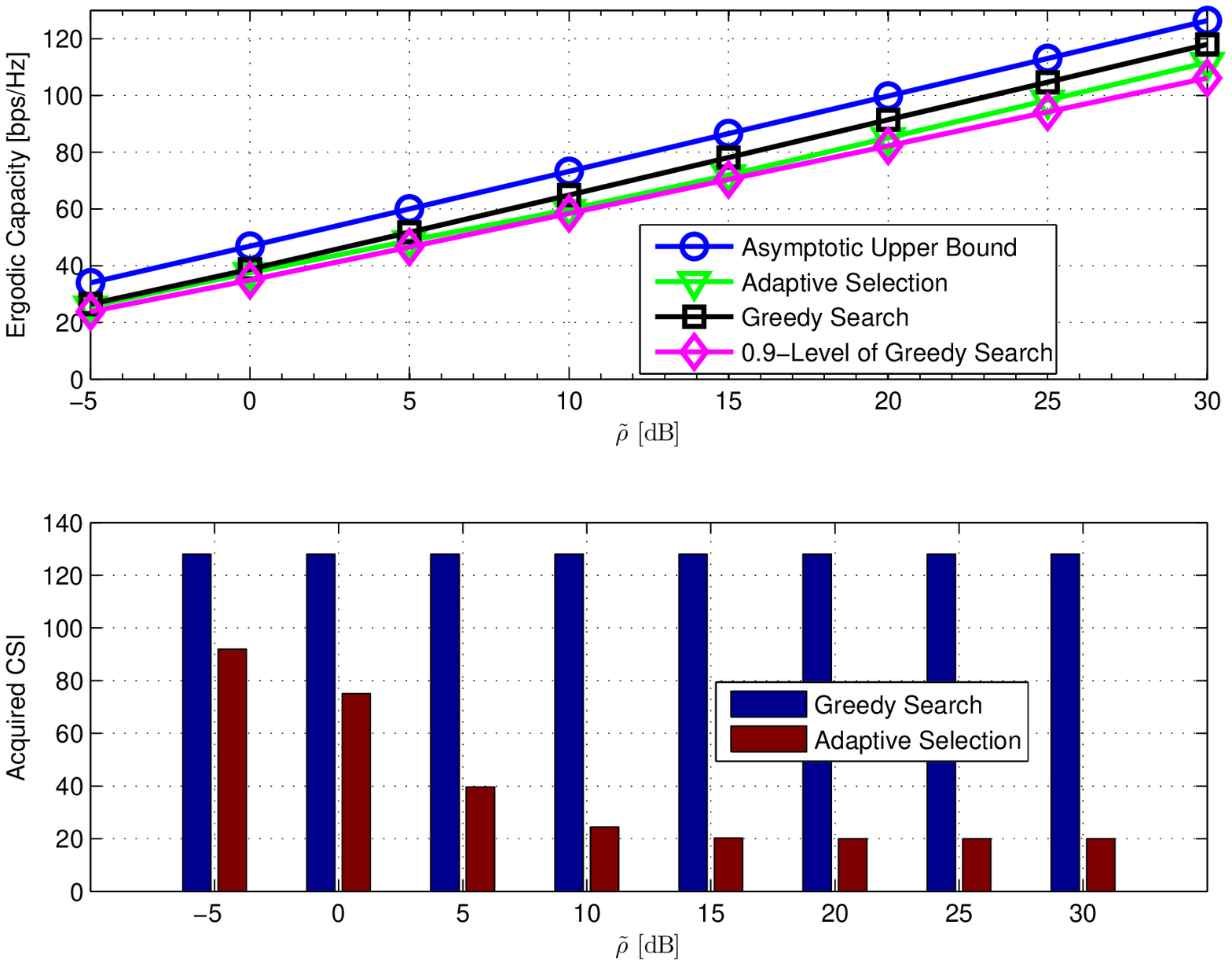}
      \label{fig11d}	        
    }
\\
    \caption{Asymptotic upper bound ($\circ$), Greedy Search based ergodic capacity($\Box$), adaptive selection based ergodic capacity ($\bigtriangledown$) and 0.9-Level ($\Diamond$) of Greedy Search based ergodic capacity versus $\tilde{\rho}$ in different antenna deployment styles are illustrated at the top of each sub-figure. The mean amount of acquired CSI verus $\tilde{\rho}$ during the whole procedure for both greedy search and adaptive search are plotted at the bottom of each sub-figure, respectively.}
    \label{figure11}
	\vspace{-0.1in}
\end{figure*}

Next, simulation results will be provided to show the advantages of the adaptive search when $L>N_{\rm{t}}$. {\figurename} \ref{figure11} compares the ergodic capacity and the mean number of acquired CSI of the proposed adaptive algorithm and greedy search in different scenarios. In each scenario, the predetermined goal is set to be the 0.85-Level of the asymptotic upper bound, i.e., $C_{\rm{final}}=0.85\mu_x$, where $\mu_x$ is the mean of asymptotic upper bound in \eqref{mx}. Moreover, the cardinality of ${\mathcal{I}}_{\rm{n}}$ is set to be 4. It can be seen from the figure that the achievable transmission rate of the adaptive search can reach the 0.9-Level of the greedy search, the reason for which is that the asymptotic upper bound is higher than the real ergodic capacity.\footnote{It should be noticed that this is only an experimental result without any quantitative theoretical bases. We do not mean that $C_{\rm{final}}$ should be set to $0.85\mu_x$ if the final goal is to reach the 0.9-Level of greedy search. The relationship between the upper bound and the achievable capacity by greedy search is still unclear.} Furthermore, the number of acquired CSI in the adaptive search is much smaller than that of the greedy search. As can be seen from the figure, there is a decreasing trend of $\Upsilon_{\rm{A}}$ as $\tilde{\rho}$ rises up, which is consist with the observed Pareto principle.

\begin{figure}[!t] 
\setlength{\abovecaptionskip}{0pt} 
\centering 
\includegraphics[width=0.45\textwidth]{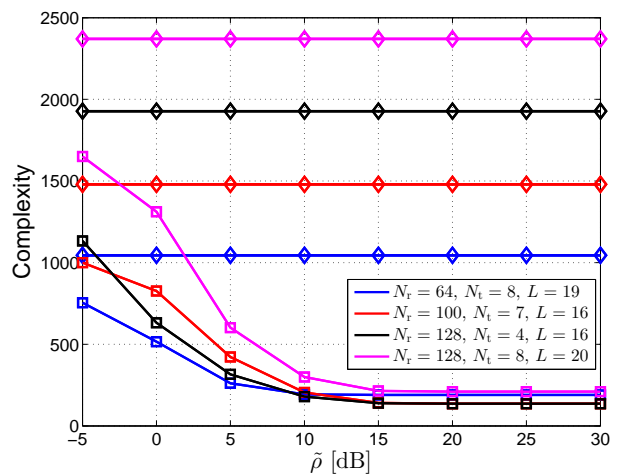} 
\caption{Complexity versus $\tilde{\rho}$ for both greedy search ($\Diamond$) and adaptive search ($\Box$) in different antenna deployment styles.}
\label{figure12}
\end{figure}

Now, let us turn to the computation complexity of the proposed algorithm. {\figurename} \ref{figure12} provides the complexity of the greedy search (full CSI is utilized) and the adaptive search (partial CSI is utilized). For the adaptive search, the predetermined goal is still set as $C_{\rm{final}}=0.85\mu_x$ and the number of acquired CSI in each step is $|{\mathcal{I}}_{\rm{n}}|=4$. The complexity of the greedy search is determined by the number of the visited nodes in the whole procedure. As shown in {\figurename} \ref{figure12}, the complexity of the adaptive search is much lower than the greedy search with full CSI especially when $\tilde{\rho}$ is high. When $\tilde{\rho}>15~\rm{dB}$, the complexity of the adaptive search is reduced by 2 orders of magnitude compared to the greedy search in each antenna deployment style.

Taken together, these results suggest that the proposed algorithm has a superior performance than the state-of-art methods. Even though there is a small loss in the achievable transmission rate, this proposed adaptive algorithm can greatly alleviate the requirement on time used for CSI acquisition. Furthermore, the complexity is much lower compared to most of the antenna selection algorithms.

\section{Conclusion}
\label{sec6}
This study set out to discuss the receive antenna selection in massive MIMO systems. A low-complexity upper capacity bound is derived based on asymptotic theory in i.i.d. Rayleigh flat fading channels. Even though the total number of the receive antennas is assumed to be infinity during the asymptotic derivation, numerical simulation indicates that the approximation is still applicable to the finite-dimensional scenarios. Furthermore, the experiments and comparison results show that the asymptotic approximated upper bound is relatively tight in both MUB and BUB cases, which means the derived result can serve as a evaluation criteria for antenna selection in massive MIMO systems. Moreover, we find that the relationship between the achievable transmission rate and the number of acquired CSI approximately follows the Pareto principle. On the basis of the observed law, an adaptive antenna selection algorithm is formulated. The proposed algorithm, which adaptively acquires the CSI and utilize it to select antennas, rquires only partial CSI and much lower computation complexity with the guarantee of considerable achievable channel capacity. These results have demonstrated the superior performance of the proposed adaptive algorithm over state-of-the-art antenna selection methods.

\vspace{12pt}
\end{document}